\documentclass[twocolumn,showpacs,aps,prb,floatfix]{revtex4}
\usepackage{graphics}
\usepackage{subfigure}
\usepackage{epsfig}

\makeatletter

\usepackage{verbatim}

\usepackage{booktabs}
\begin{document}
\makeatletter

\input epsf
\def\prb{Phys. Rev. B}
\def\prl{Phys. Rev. Lett.}
\def\pla{Phys. Lett. A}
\def\pr{Phys. Rev.}
\def\be{\begin{equation}}
\def\ee{\end{equation}}
\def\ba{\begin{eqnarray}}
\def\ea{\end{eqnarray}}

\makeatother

\title{Magnetic Excitations of Stripes and Checkerboards in the Cuprates}

\author{D.~X.~Yao$^1$, E.~W.~Carlson$^2$, and D.~K.~Campbell$^1$}

\affiliation{
(1) Department of Physics and Department of Electrical and Computer Engineering, Boston University, Boston, Massachusetts 02215, USA \\
(2) Department of Physics, Purdue University, West Lafayette, Indiana 47907, USA 
}

\date{July 8, 2006}

\begin{abstract}
We discuss the magnetic excitations of
well-ordered stripe and checkerboard phases, including the high energy magnetic excitations of
recent interest and possible connections to the ``resonance peak'' in cuprate
superconductors. Using a suitably parametrized Heisenberg model  and spin
wave theory, we study a variety of magnetically ordered 
configurations, including vertical and  diagonal site- and bond-centered stripes and simple checkerboards.
We calculate the expected neutron scattering intensities as a function of
energy and momentum.  
At zero frequency, the satellite peaks of 
even square-wave stripes are suppressed  by as much as a factor of $34$
below the intensity of the main incommensurate peaks.  
We further find that at low energy, spin wave cones may  not always be resolvable
experimentally.
Rather, the intensity as a function of position around the cone depends strongly on the
coupling across the stripe domain walls.  At intermediate energy, we find a saddlepoint
at $(\pi,\pi)$ for a range of couplings, and discuss its possible connection
to the ``resonance peak" observed in neutron scattering experiments on cuprate superconductors.  At high energy,
various structures are possible as a function of coupling strength and configuration,
including a high energy square-shaped continuum originally attributed to the
quantum excitations of spin ladders.  
On the other hand, we find that simple checkerboard patterns are inconsistent with experimental results from neutron scattering.
\end{abstract}
\pacs{74.72.-h, 74.25.Ha, 75.30.Ds, 76.50.+g}

\maketitle

%%%%%%%%%%%%%%%%%%%%%%%%%%%%%%%%%%
\section{Introduction}
%%%%%%%%%%%%%%%%%%%%%%%%%%%%%%%%%%

Strong correlations in electronic systems, and especially competing interactions,
can cause mesoscale electronic structure to spontaneously develop.  
In cuprate superconductors and the related nickelate compounds,
several locally inhomogenous electronic phases have been proposed,
involving charge order, spin order, or both.  Several experimental probes
corroborate some level of local order, including STM,\cite{checknccoc,hoffman,aharon}
neturon scattering,\cite{mookchg,tranquada04a,hayden04,aeppli}, NMR\cite{hammelwipeout}
and $\mu$SR studies.\cite{neidermeyer}
Recent experimental advances have made possible the detection of high energy
neutron scattering spectra.\cite{hayden04,tranquada04a,buyers04}  There has been much interest in the interpretation of these spectra in the cuprates, especially since the high energy results
%EC updated this to say LBCO, not LSCO. 
from neutron scattering
on La$_{2-x}$Ba$_x$CuO$_4$ (LBCO) and YBa$_2$Cu$_3$O$_{6+\delta}$ (YBCO)
exhibit universal behavior.\cite{tranquada04a,hayden04}
This paper extends the previous work of Refs.\cite{spinprb1,spinprl} to look at a broader range of
ordered structures, and also to explore the scattering patterns which are possible at high 
energies.  It has been suggested, for example, that the high energy magnetic excitations
in LBCO and YBCO
may be due to the quantum excitations of stripes.\cite{vojta-ulbricht,uhrig04a,tranquada04a,seibold04}
We have reported elsewhere\cite{spinprl} that high energy excitations
near a quantum critical point (QCP) to disordered ladders\cite{vojta-ulbricht,uhrig04a,seibold04,vojta-sachdev} 
can strongly resemble semiclassical excitations, due to the small critical exponent $\eta = 0.037$ 
associated with this QCP.  

One continuing mystery about the low energy results has been 
the lack of observed spin satellite peaks in neutron scattering,
and also that spin wave cones are rarely observed in the cuprates. Rather, what is often seen in the
low energy regime may be more accurately termed ``legs of scattering''. 
Both of these results have raised questions about a ``stripe" interpretation
of the data.  We show below that at $\omega = 0$, satellite peaks for 
even the most extreme case of square-wave spin
stripes have very low intensity,
and may not be resolvable without very high experimental resolution.  
In addition, although a spin ordered state results in spin wave cones due to Goldstone's theorem,
the intensity is not always uniform.  Rather, the intensity can be gathered on the inner branch of the spin wave cones [the side nearest $(\pi,\pi)$], 
or on the outer branch, depending upon the relative strength of the spin
coupling across the charge stripes.  
For this reason, while spin wave cones are always present  for ordered
spin stripes, they may not yet be resolvable experimentally.  

An important point we wish to emphasize is that although
stripes are a unidirectional modulation in an otherwise antiferromagnetic
texture, they are a fully two-dimensional (2D) spin order, with a 2D magnetic 
Bravais lattice, which gives rise to 2D scattering signals at all energies. 
Our results never show streaks of scattering in the low energy structure, although
streaks are possible in the high energy structure for weak coupling, as we report below.

%%%%%%%%%%%%%%%%%%%%%%%%%%%%%%%%%%
\section{Model and Method}
%%%%%%%%%%%%%%%%%%%%%%%%%%%%%%%%%%

%%%%%%%%%%%%%% FIGURES:  V and D Stripes  %%%%%%%%%%%%%%%%%%%%
\begin{figure}[Htb]
{\centering
  \subfigure[VS4]{\resizebox*{0.43\columnwidth}{!}{\includegraphics{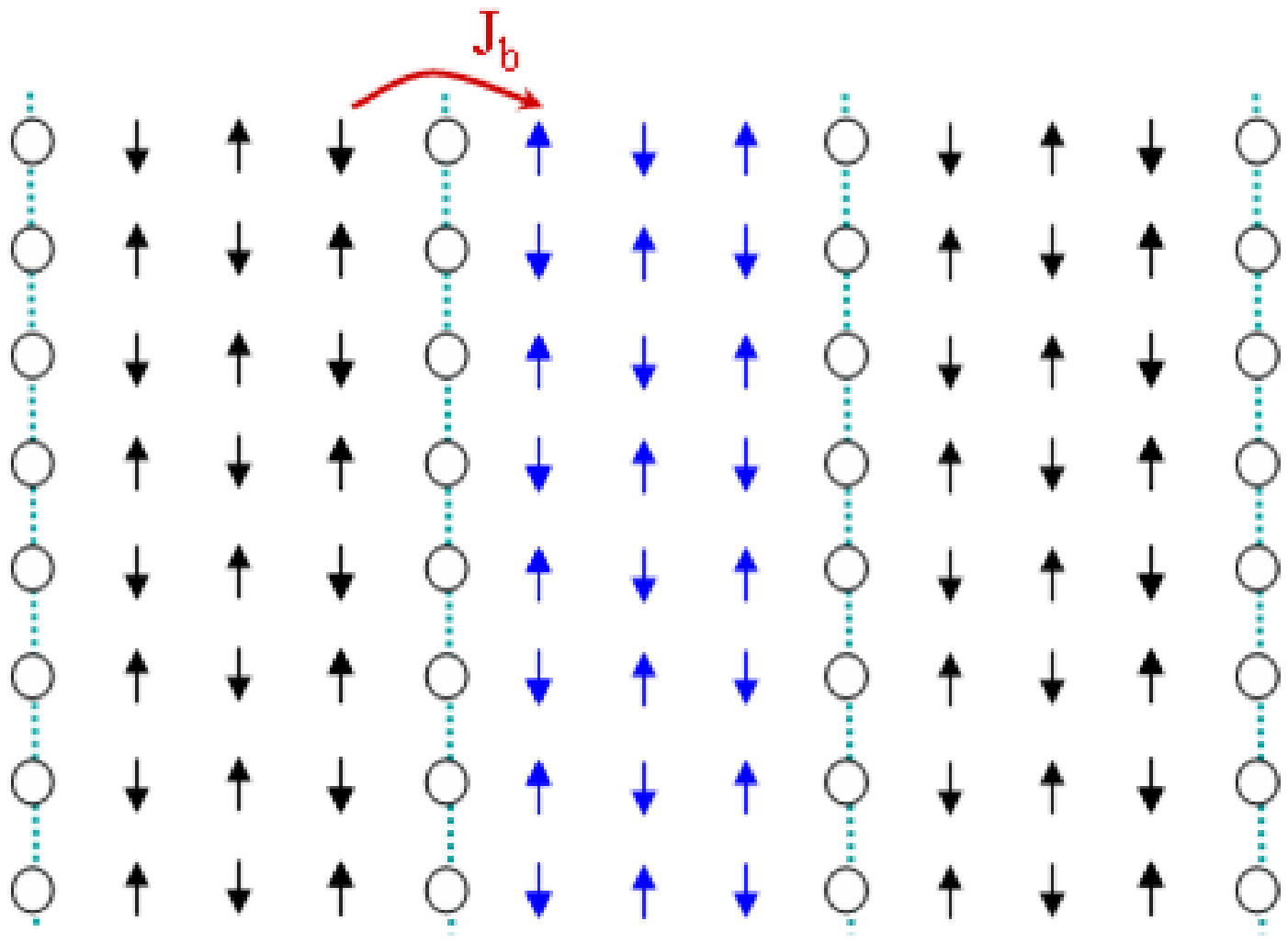}}}
  \hspace{0.05\columnwidth}
  \subfigure[VB4]{\resizebox*{0.43\columnwidth}{!}{\includegraphics{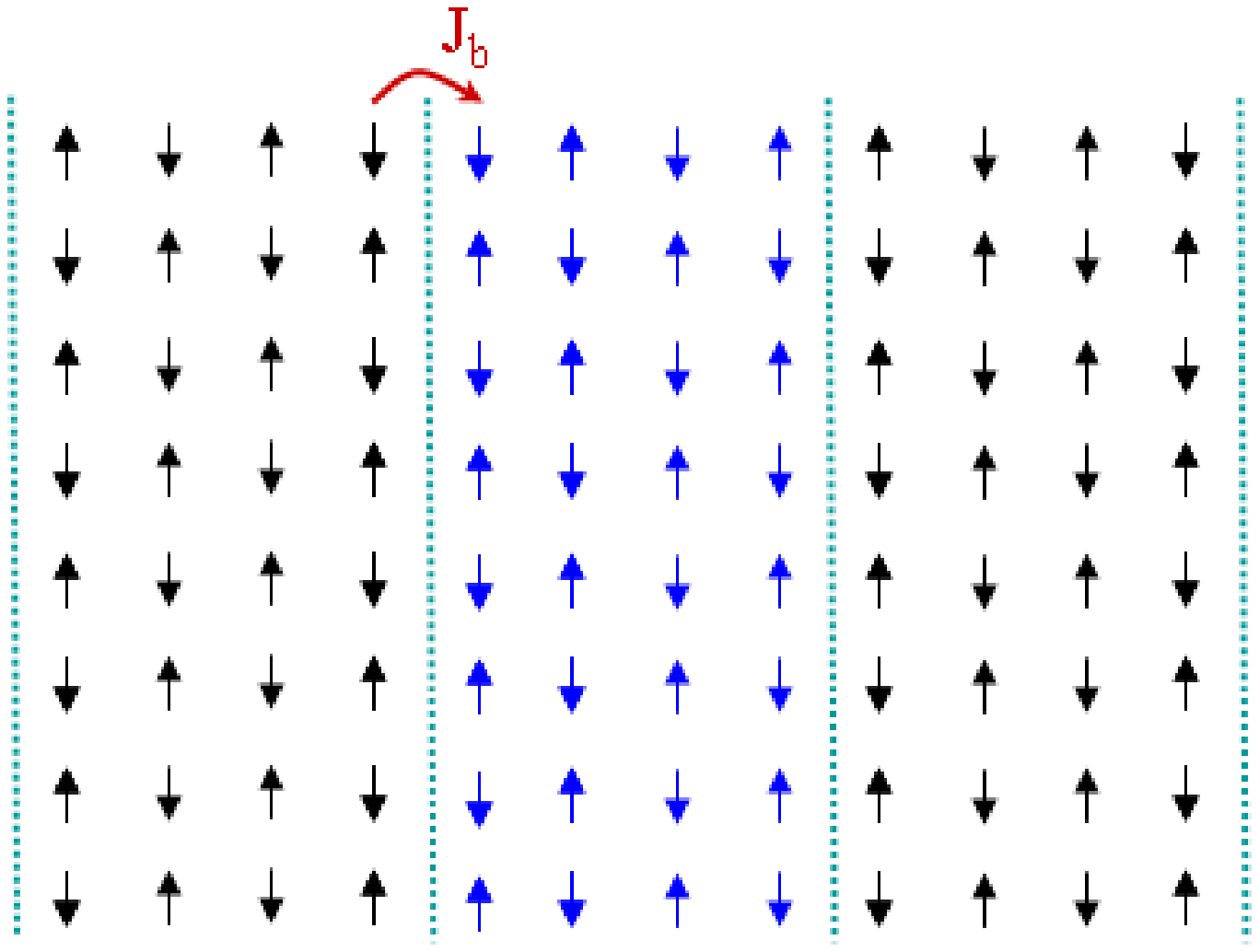}}} \par}
{\centering  
  \subfigure[DS3]{\resizebox*{0.4\columnwidth}{!}{\includegraphics{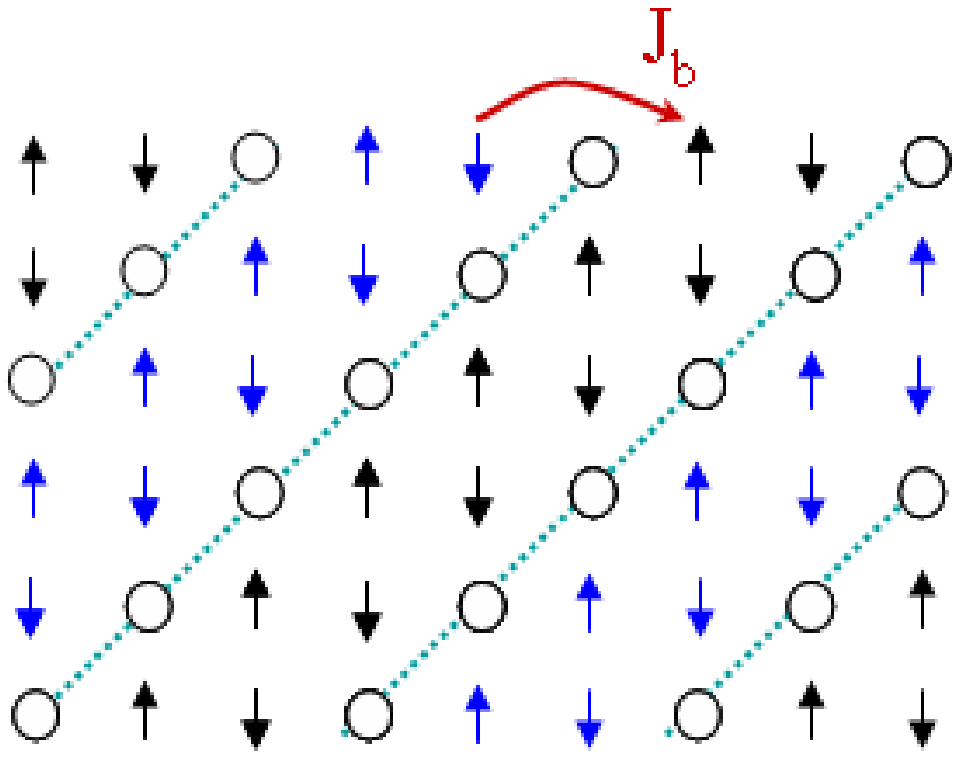}}}
  \hspace{0.05\columnwidth}
  \subfigure[DB2]{\resizebox*{0.4\columnwidth}{!}{\includegraphics{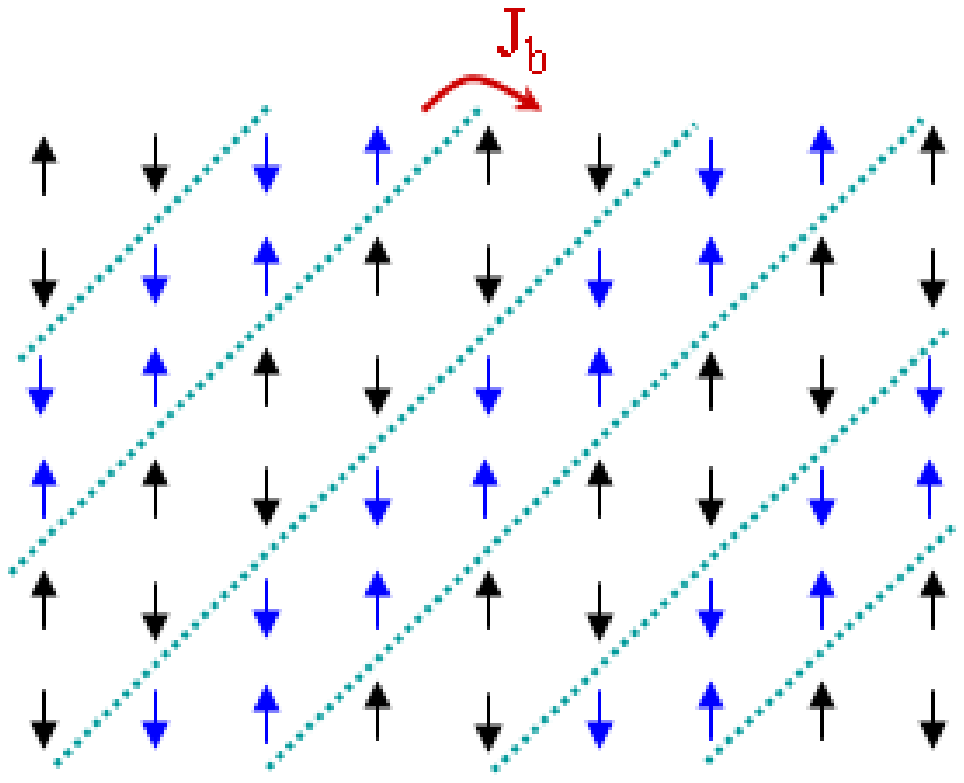}}}   \par}
\caption{(Color online) Stripe configurations studied in this paper.  The coupling between nearest neighbor
spins is $J_a$, and the coupling between spins across a domain wall is $J_b$,
as indicated in the figures.
(a) VS4: Vertical site-centered stripes with spacing $d=4$. (b) VB4: Vertical 
bond-centered stripes with spacing $d=4$.  (c) DS3: Diagonal site-centered stripes with spacing
$d=3$.  (d) DB2:  Diagonal bond-centered stripes with spacing $d=2$.  
%\label{lattice}
\label{fig:allstripes}}
\end{figure}
%%%%%%%%%%%%%% FIGURES:  C Configurations %%%%%%%%%%%%%%%%%%%%
\begin{figure}[Htb]
{\centering
  \subfigure[CS3]{\resizebox*{0.4\columnwidth}{!}{\includegraphics{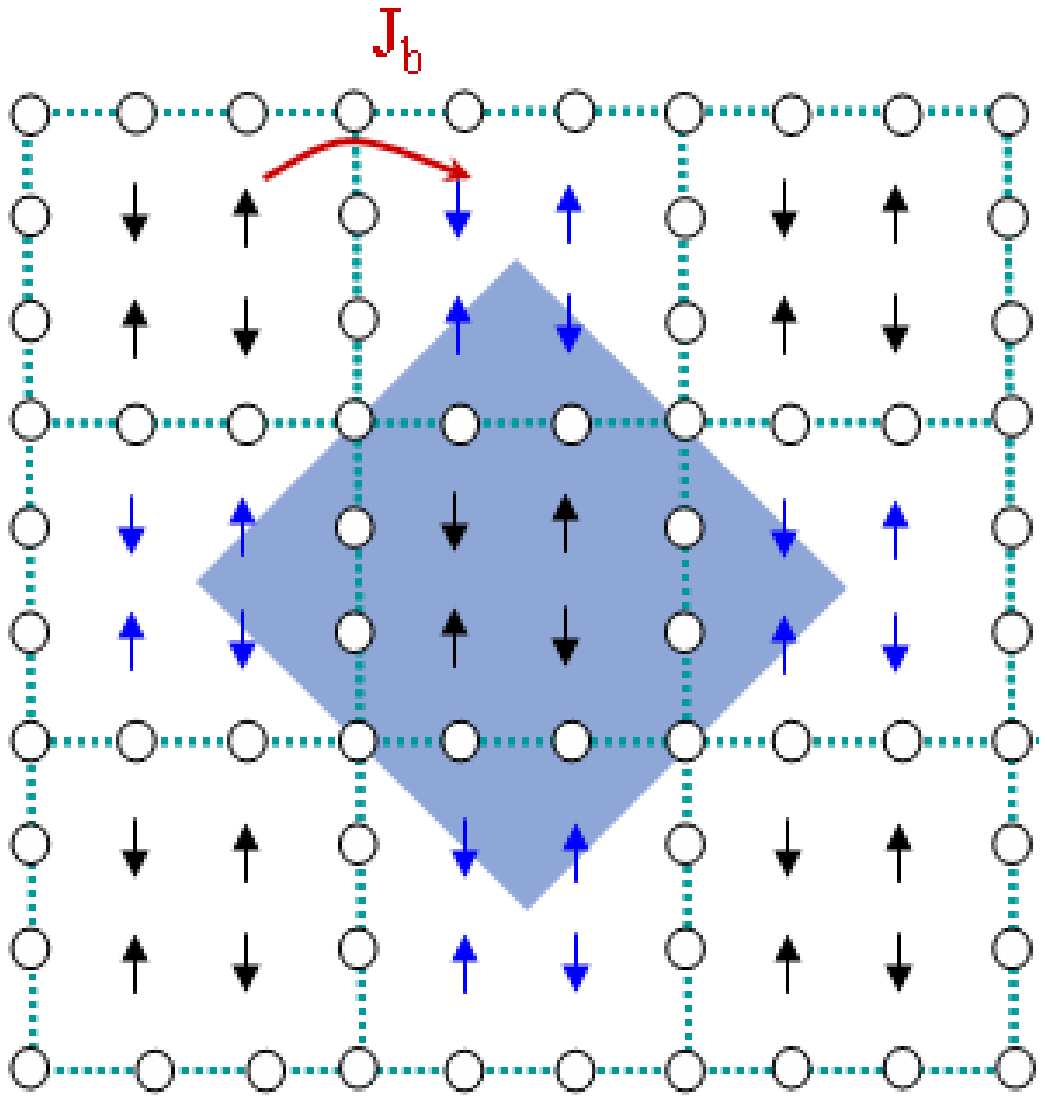}}}
  \hspace{0.05\columnwidth}
  \subfigure[CS4]{\resizebox*{0.4\columnwidth}{!}{\includegraphics{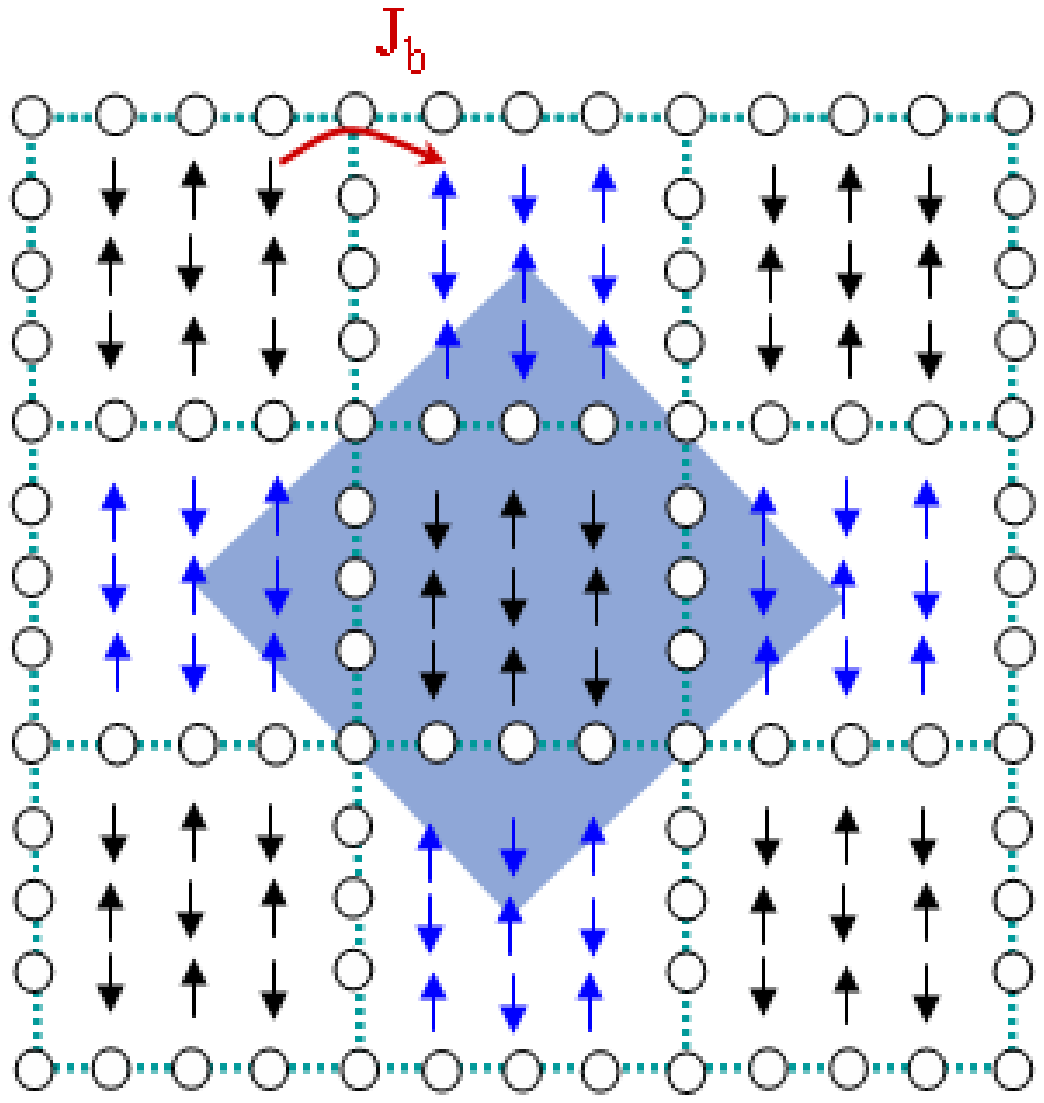}}} \par}
{\centering
  \subfigure[CB2]{\resizebox*{0.45\columnwidth}{!}{\includegraphics{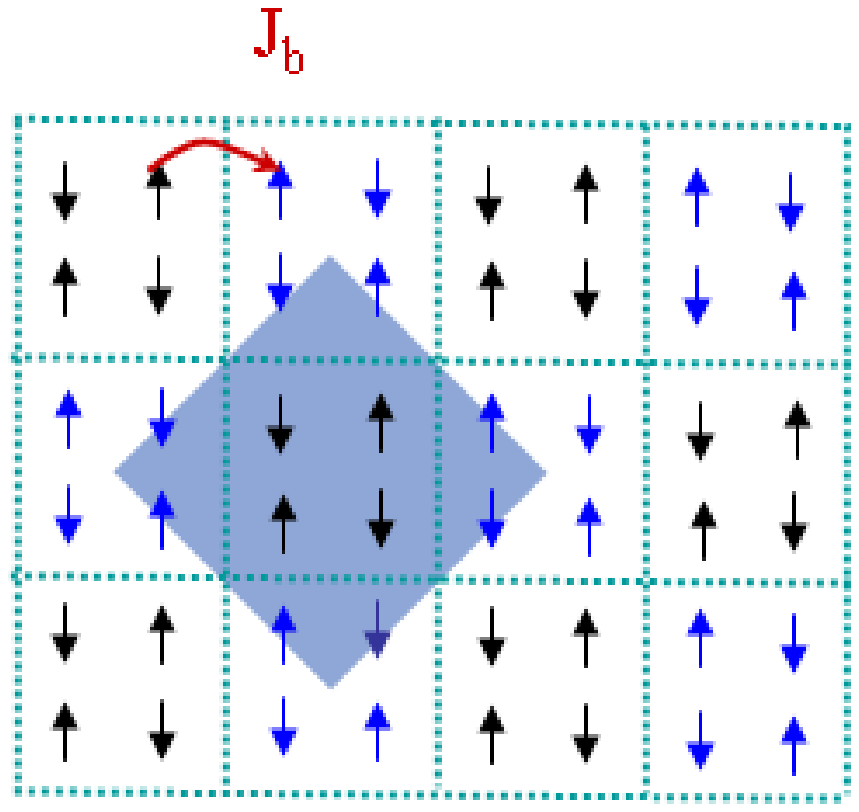}}}
  \hspace{0.05\columnwidth}
  \subfigure[CB3]{\resizebox*{0.40\columnwidth}{!}{\includegraphics{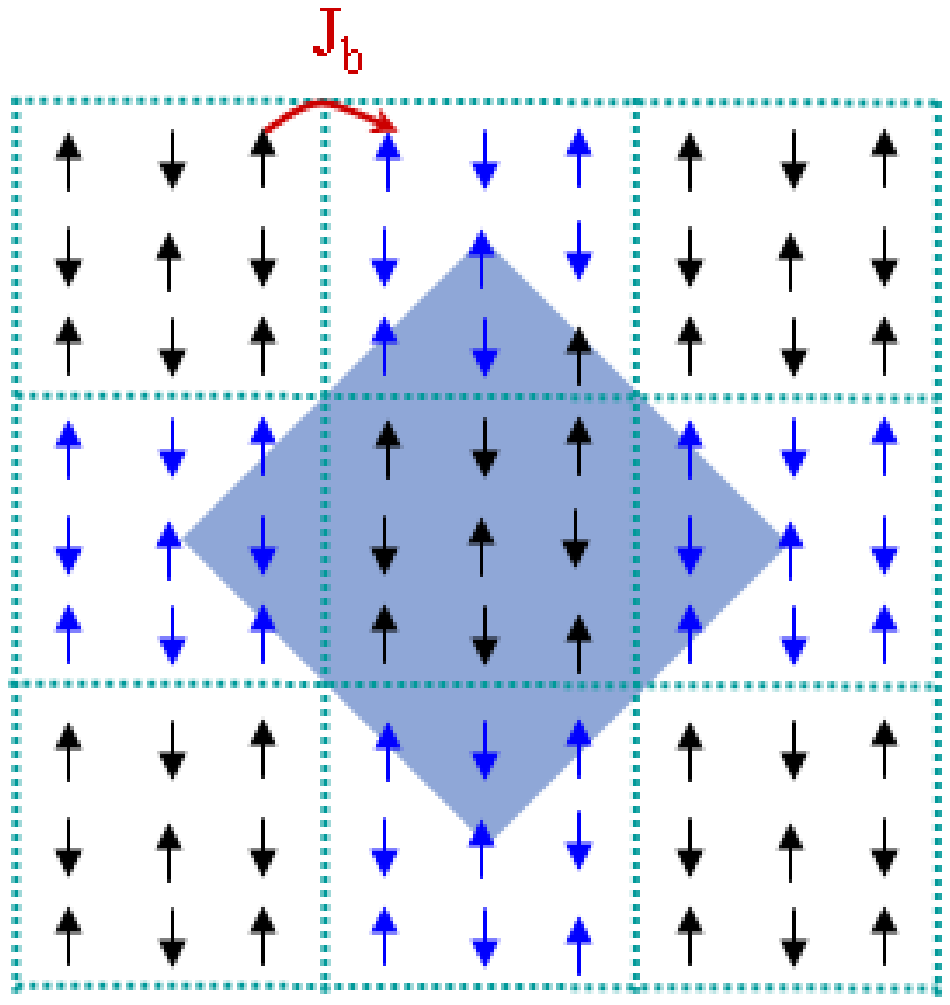}}} \par}
\caption{(Color online) Checkerboard patterns.  The coupling between nearest neighbor
spins is $J_a$, and the coupling between spins across a domain wall is $J_b$,
as indicated in the figures.
(a) CS3: Checkerboard sited-centered pattern with  spacing $d=3$.  (b) CS4: Checkerboard site-centered pattern with 
  spacing $d=4$.
(c) CB2: Checkerboard bond-centered pattern with  spacing $d=2$. (d)CB3: Checkerboard bond-centered pattern with spacing $d=3$.}
\label{fig:checkerbond}
\end{figure}
%%%%%%%%%%%%%%%%%%%%%%%%%%%%%%%%%%

In this work we concentrate on static stripes and checkerboards as arrays of 
antiphase domain walls in an otherwise
antiferromagnetic texture. We are interested solely in the response
of the spin degrees of freedom.  
The dynamics of the charge component which must reside on every domain wall is to 
renormalize the effective spin couplings in the Hamiltonian.
We thus use a suitably parametrized Heisenberg model to describe the 
elementary excitations of spin stripes, employing a 
modulation of the exchange integral 
to capture the effective spin coupling in a well ordered stripe or checkerboard phase.
\begin{equation}
H= \frac{1}{2} \sum_{\mathbf{r},\mathbf{r'}} J_{\mathbf{r},\mathbf{r'}} \mathbf{S}_{\mathbf{r}} \mathbf{S}_{\mathbf{r'}},
\label{model}
\end{equation}
where \(J_{\mathbf{r},\mathbf{r'}}\) is the effective spin coupling.  
We work in units where $\hbar = 1$.
Nearest neighbor couplings
are antiferromagnetic with $J_{\mathbf{r},\mathbf{r'}}=J_a>0$ within each antiferromagnetic
patch. Couplings across a domain wall are different, and depend upon the configuration (such as 
spacing and direction of the domain walls).  These are enumerated below.  
When comparing to, {\em e.g.}, the cuprates (nickelates),
our lattice corresponds to the copper (nickel)  sites within the copper-oxygen (nickel-oxygen) planes.
With these materials in mind, we restrict ourselves to patterns embedded in a two-dimensional square lattice.

%%%%%%%%%%%%%%%%%%%%%%%%%%%%%%%%%%
\subsection{Stripe and Checkerboard Configurations}
%%%%%%%%%%%%%%%%%%%%%%%%%%%%%%%%%%
In the presence of a host crystal, stripes are constrained by
the symmetry of the crystal to 
run along major crystallographic 
directions.  We consider two classes of stripe patterns, vertical and diagonal,
as well as checkerboard patterns.  
These classifications concern the pattern of the antiphase domain walls
in the antiferromagnetism.  
We use the term ``vertical'' to describe stripes whose domain walls run parallel
to the Cu-O (Ni-O) bond direction, and the term ``diagonal'' to
describe stripes that run $45^o$ from that direction.
These  stripe patterns are depicted in
Fig.~\ref{fig:allstripes}.

A further distinction between types of stripes and checkerboards
depends on  where the antiphase domain walls sit with respect to the atomic lattice.
``Site-centered'' stripes have domain walls which are centered 
on the sites of the square lattice ({\em i.e.} on the
nickel or copper sites).   These have 
antiferromagnetic coupling
$J_{\mathbf{r},\mathbf{r'}}=J_b>0$ across the domain wall. 
On the other hand, when the 
domain wall is situated between two square lattice sites
({\em i.e.} between nickel or copper sites, on the planar oxygens),
the stripes are ``bond-centered''.
In this case the coupling across the domain wall is effectively ferromagnetic
(to preserve the antiphase nature of the domain walls),
and $J_b<0$.\cite{sitecentered}

%EC updated the next paragraph to include definitions of abbreviations, and also to include a couple of sentences about which material and which probe to satisfy referee 2.
Checkerboard patterns have been proposed to explain the real space 
structure observed in STM experiments on 
Bi$_2$Sr$_2$CaCu$_2$O$_{8+\delta}$\cite{hoffman} (BSCCO) and 
Ca$_{2-x}$Na$_x$CuO$_2$Cl$_2$\cite{checknccoc} (Na-CCOC).
It is important to note that STM is a surface probe, and to date, the $4\times 4$ pattern
observed in BSCCO and Na-CCOC has not been confirmed by neutron scattering or other bulk probes
in these materials.  Likewise, the $4 \times 4$ pattern has not yet been confirmed via STM to be present in the lanthanum and yttrium compounds.  
Nevertheless, it has been noted that the length scale of the charge periodicity observed in BSCCO and Na-CCOC via STM is half that of the spin periodicity found in neutron scattering in related
lanthanum-based and yttrium-based cuprate superconductors, and that therefore the two probes may be
observing similar charge and spin modulations.   
Since the neutron scattering shows satellite peaks around antiferromagnetism,
rather than a peak at the antiferromagnetic wavevector $(\pi,\pi)$,
any universal spin texture which is consistent with the proposed checkerboard
pattern must also incorporate antiphase domain walls in the corresponding
spin texture.  Representative, simple spin checkerboard configurations
are shown in Fig.~\ref{fig:checkerbond}.  The parameter $d$ is the spacing
between domain walls.
The low energy peaks of these simple checkerboard patterns
%EC added reference
are inconsistent with neutron scattering data,\cite{vojta-sachdev,andersen03} 
as we will show
in Sec.~\ref{sec:check}, although more complicated checkerboard patterns\cite{abbamonte}
may be consistent.

We use the notation VSd,VBd,DSd and DBd to refer to vertical (V)
or diagonal (D) stripes of spacing d in a site (S)- or bond (B)-centered
configuration,\cite{spinprb1} as illustrated in Fig.~\ref{fig:allstripes}.
The notation CSd and CBd refers to checkerboards of site (S)- or bond (B)-centered
domain walls which are $d$ lattice sites apart.

%%%%%%%%%%%%%%%%%%%%%%%%%%%%%%%%%%
\subsection{Spin-wave method}
%%%%%%%%%%%%%%%%%%%%%%%%%%%%%%%%%%
We use linear spin-wave theory to explore the semiclassical spinwave excitations
of well-ordered spin stripes, as modeled by Eq.~\ref{model}. Ladder operators may be 
used to rewrite the Hamiltonian in term of spin wave excitations above
the semiclassical  ground state:
\begin{equation}
H=\frac{1}{2} \sum_{<\mathbf{r},\mathbf{r'}>} J_{\mathbf{r},\mathbf{r'}}
[S^z_{\mathbf{r}}S^z_{\mathbf{r'}} + \frac{1}{2}(S^+_{\mathbf{r}}S^-_{\mathbf{r'}}+S^-_{\mathbf{r}}S^+_{\mathbf{r'}})].
\end{equation}
The spin ladder operators may be approximated as 
Holstein-Primakoff bosons, a standard procedure described
elsewhere,\cite{spinprb1,assa,kruger03} in order to calculate 
the spin wave excitation spectrum as well as the 
zero-temperature dynamical structure factor, 
\begin{equation}
S(\mathbf{k}, \omega)=\sum_f \sum_{i=x,y,z} |\left<f|S^i (\mathbf{k})|0\right>|^2 \delta (\omega-\omega_f)
\end{equation}
which is proportional  to the
expected neutron scattering intensity for single magnon excitations.
Here $\left|0 \right>$ is the magnon vacuum state and $\left|f \right>$ denotes the final state of
the spin system with excitation energy $\omega_f$.
We report single magnon excitations, and neglect possible
spin-wave interactions since they are higher order effects. 

%%%%%%%%%%%%%%%%%%%%%%%%%%%%%%%%%%
\section{Results: Spectra of Vertical Stripes \label{results:vert}}
%%%%%%%%%%%%%%%%%%%%%%%%%%%%%%%%%%

%%%%%%%%%%% FIGURE:  VS4 Jb = 0.4 Ja  %%%%%%%%%%%%%
\begin{figure}
\resizebox*{0.95\columnwidth}{!}{\includegraphics{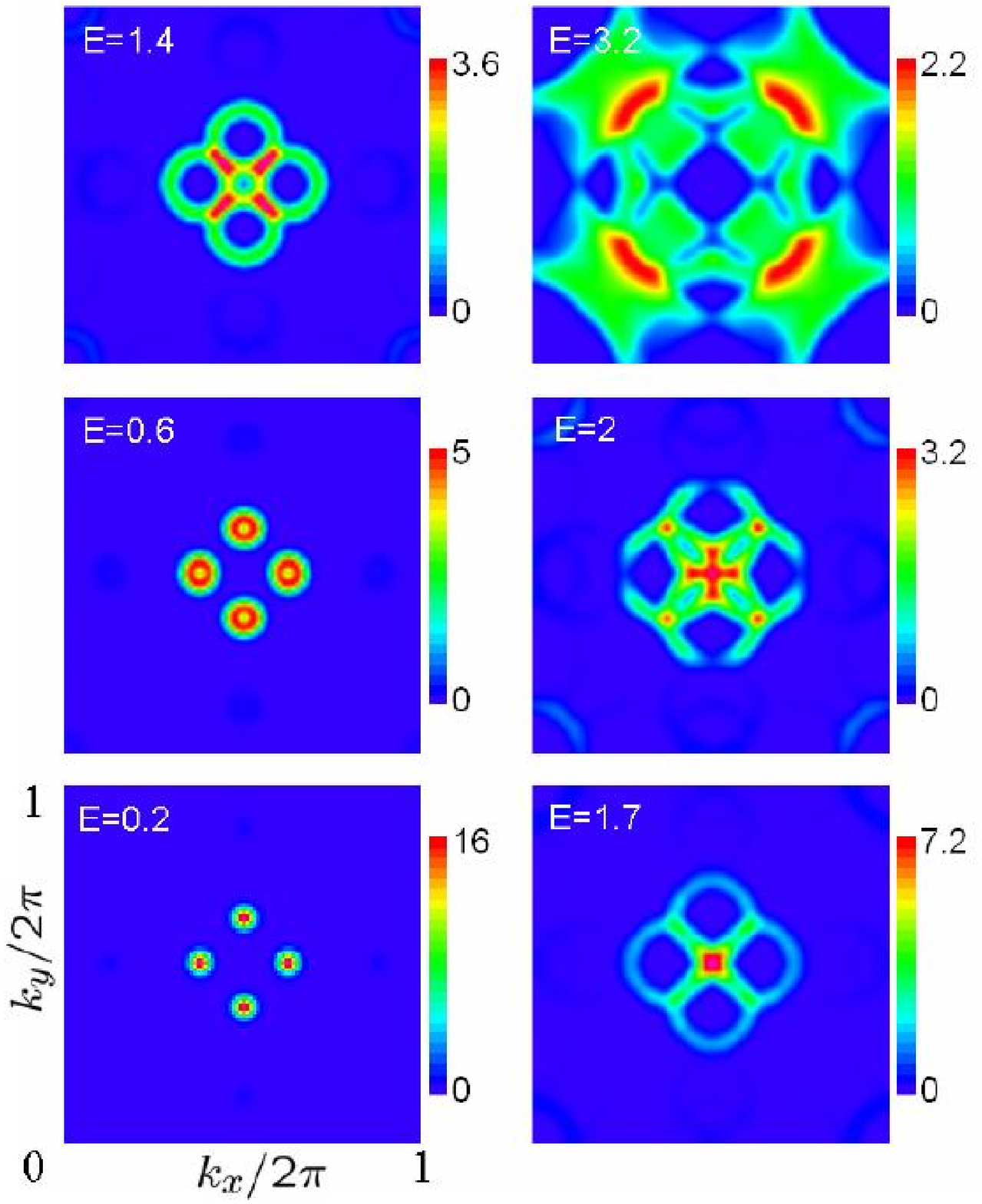}}
\caption{(Color online) VS4 at $J_b = 0.4 J_a$: Constant energy cuts with windows $0.2Ja$ 
for twinned vertical, site-centered stripes of spacing $d=4$ at $J_b =  0.4 J_a$.
The energy $E$ is in units of $J_a S$.\label{fig.vs4_0.4}}
\end{figure}
%%%%%%%%%%% FIGURE:  VS4 Jb = Ja  %%%%%%%%%%%%%
\begin{figure}
%\resizebox*{1\columnwidth}{!}{\includegraphics{vs4.eps}}
\resizebox*{0.95\columnwidth}{!}{\includegraphics{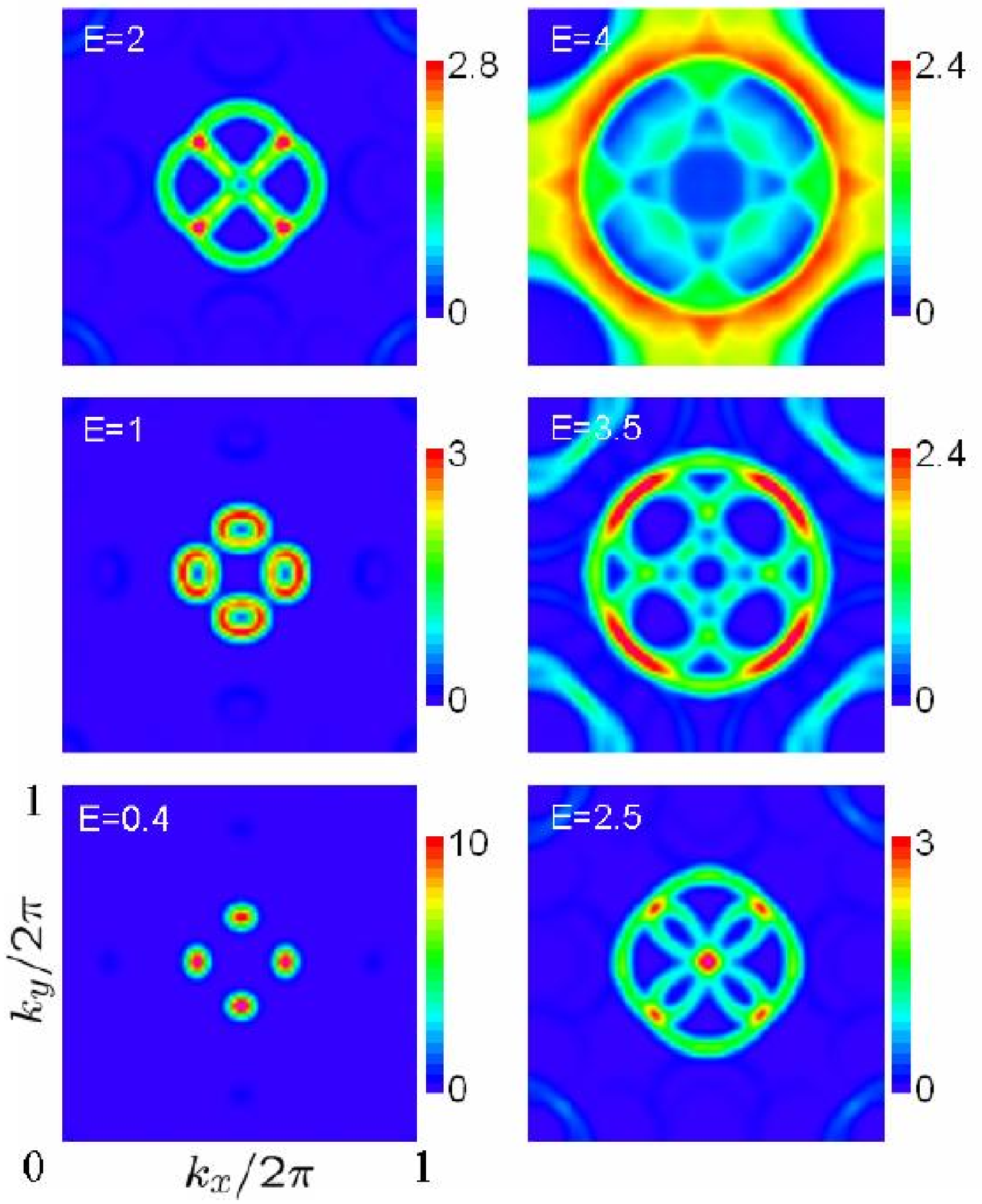}}
\caption{(Color online) VS4 at $J_b =  J_a$: Constant energy cuts with windows $0.2Ja$ 
for twinned vertical, site-centered stripes of spacing $d=4$ at $J_b =  J_a$.
The energy $E$ is in units of $J_a S$.\label{fig.vs4_1.0}}
\end{figure}
%%%%%%%%%%% FIGURE:  VB4 Jb = 0.4 Ja %%%%%%%%%%%
\begin{figure}
\resizebox*{0.95\columnwidth}{!}{\includegraphics{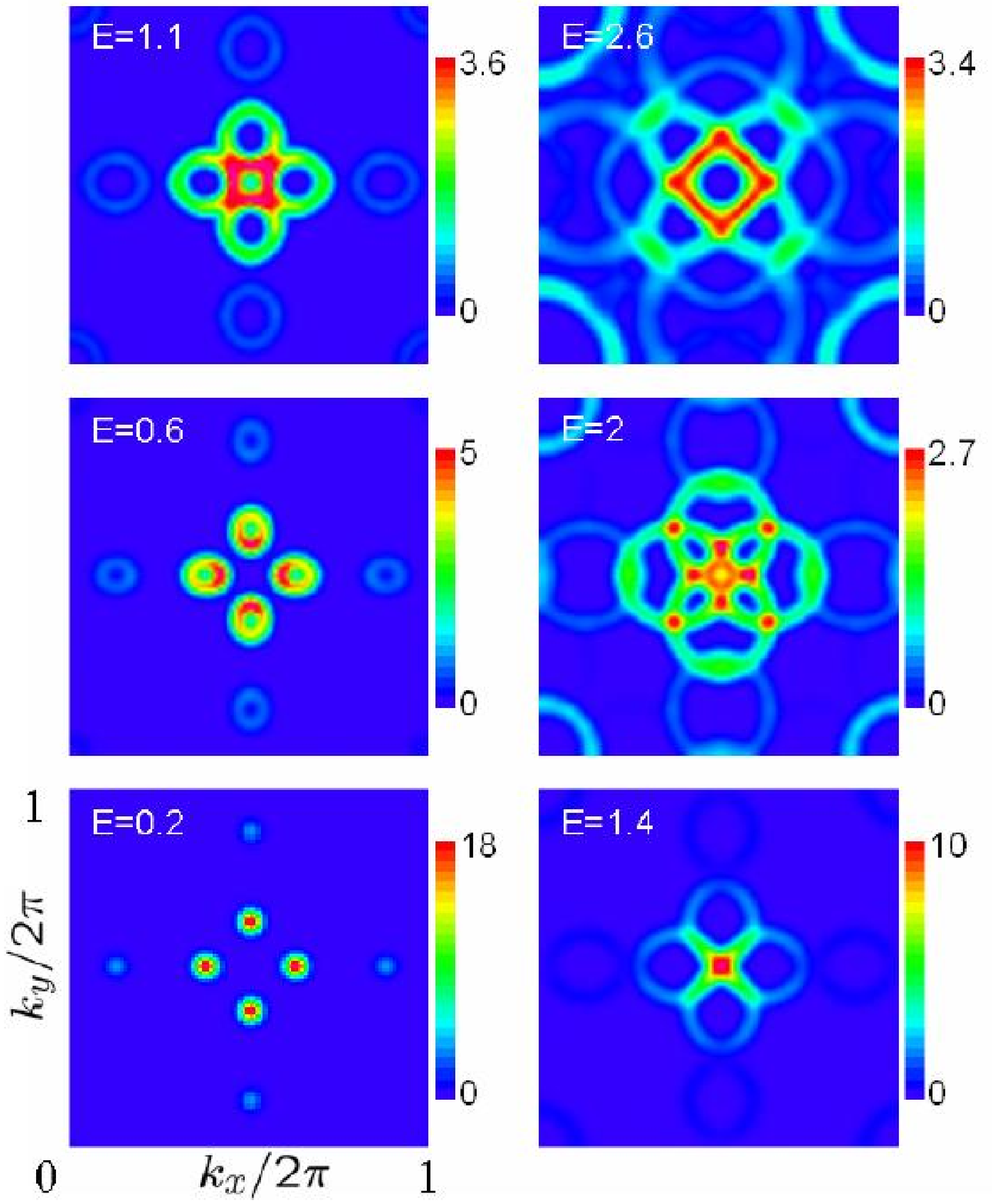}}
\caption{(Color online) VB4 at $J_b=-0.4J_a$: Constant energy cuts with windows $0.2Ja$ for twinned vertical,
  bond-centered stripes of spacing $d=4$ at $J_b=-0.4J_a$. The energy $E$ is in units of $J_a S$.
\label{fig.vb4_0.4}}
\end{figure}
%%%%%%%%%%% FIGURE:  VB4 Jb = 1.5 Ja %%%%%%%%%%%
\begin{figure}
\resizebox*{0.95\columnwidth}{!}{\includegraphics{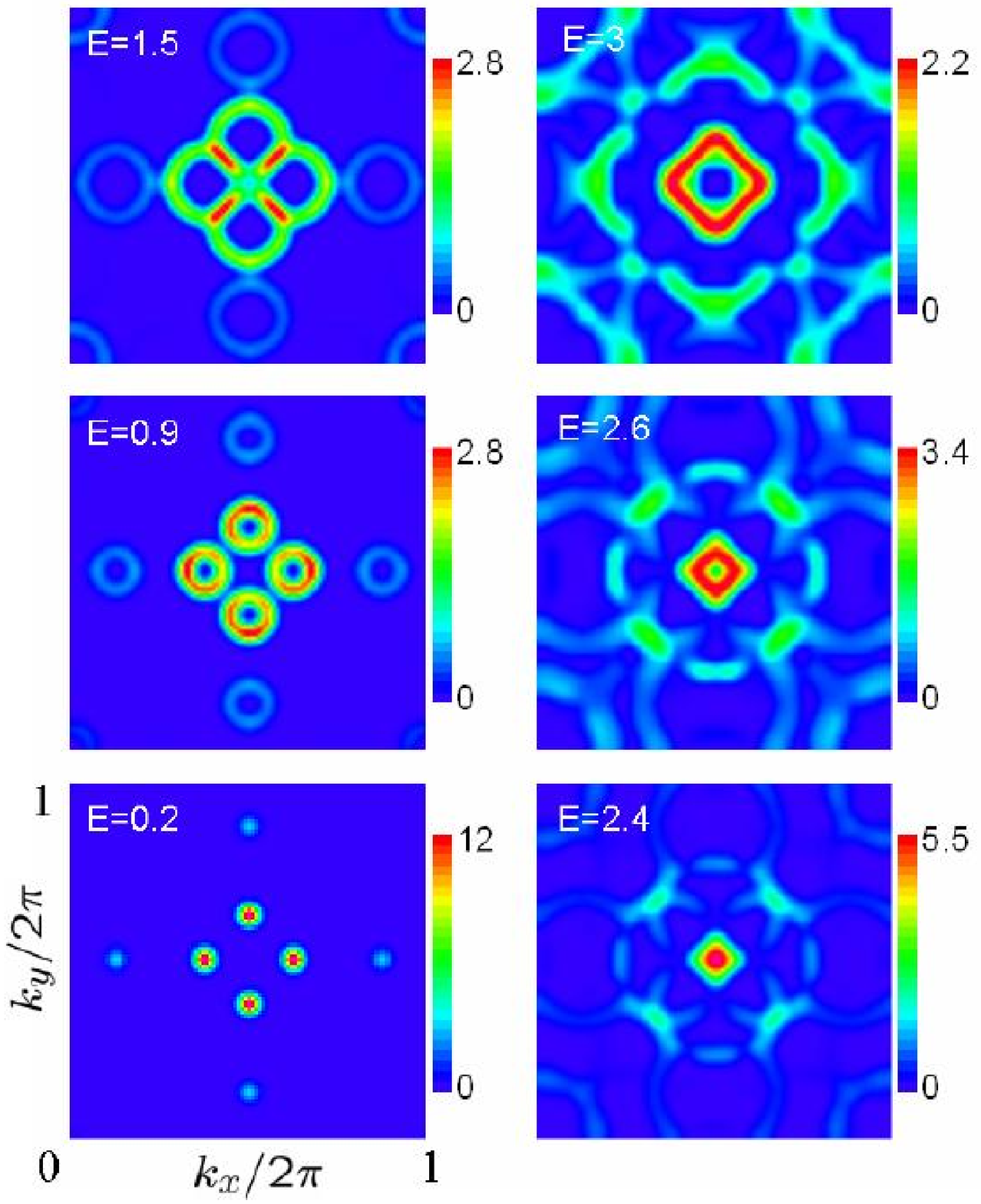}}
\caption{(Color online) VB4 at $J_b=-1.5J_a$: Constant energy cuts with windows $0.2Ja$ for twinned vertical,
  bond-centered stripes of spacing $d=4$ at $J_b=-1.5J_a$. The energy $E$ is in units of $J_a S$.
\label{fig.vb4_1.5}}
\end{figure}
%%%%%%%%%%%%%% FIGURE: Intensity Ratio %%%%%%%%%%%%%%%%%%%%
\begin{figure}[Htb]
{\centering
  \subfigure[VS4 Intensity Ratio]{\resizebox*{0.8\columnwidth}{!}{\includegraphics{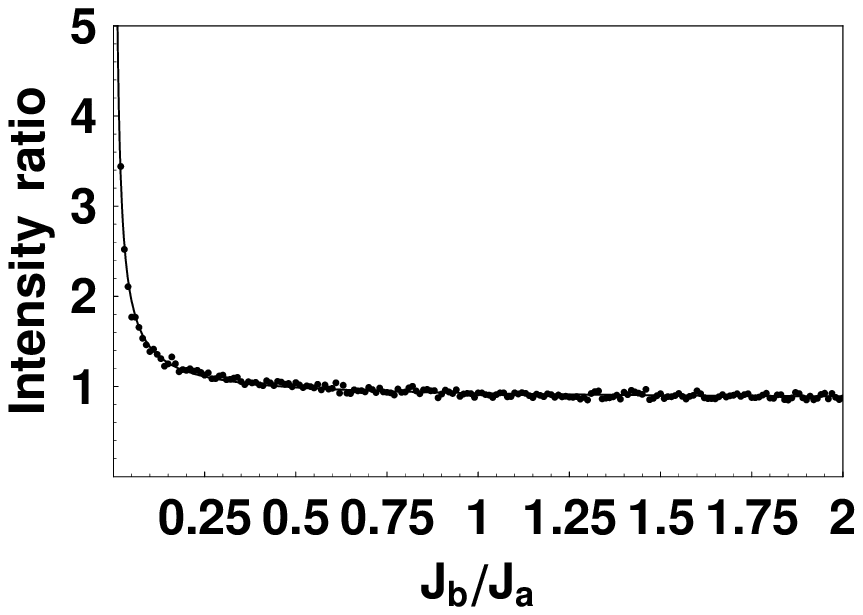}}}
  \subfigure[VB4 Intensity Ratio]{\resizebox*{0.8\columnwidth}{!}{\includegraphics{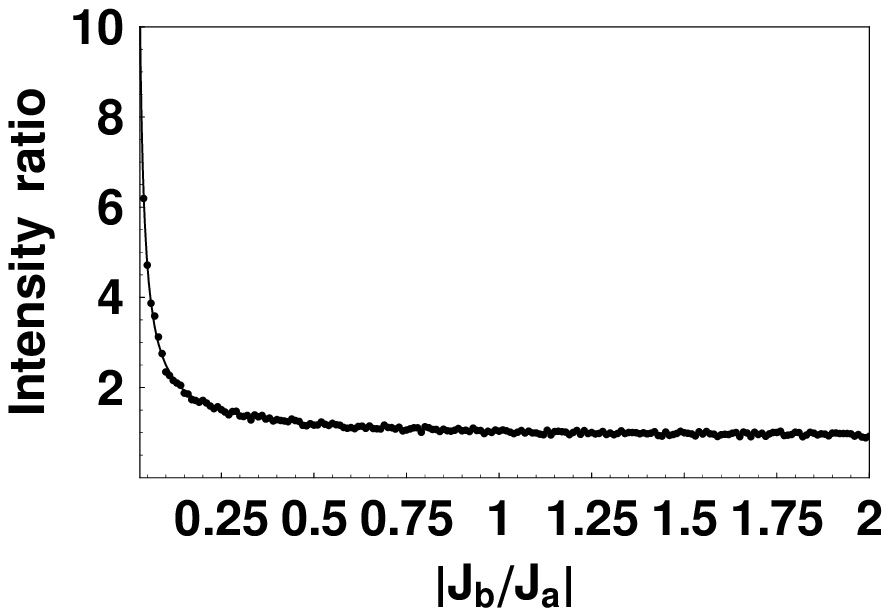}}} \par}
\caption{Intensity ratio between the ``inner" and ``outer" branches of the spin wave
cones as a function of coupling ratio $|J_b/J_a|$ 
for vertical stripes.  The ratio is calculated
at constant low energy $E=0.4J_aS$.  The solid points are our numerical calculations of
the intensity ratio.  The solid line is a fit to the results, as explained
in the text.
(a) VS4:  Vertical, site-centered stripes of spacing $d=4$.
(b) VB4:  Vertical, bond-centered stripes of spacing $d=4$.}
\label{fig:intensityratio}
\end{figure}
%%%%%%%%%%%%%%%%%%%%%%%%%%%%%%%%%%%%%%%%

In this section, we report new results on  the expected neutron scattering intensity
at constant energy  for vertical and diagonal stripes.  
In order to make a comparison with the experiments, we report constant energy
plots of intensity in the $(k_x,k_y)$ plane.
We work in tetragonal units, where the $k_x$ and $k_y$ directions
are oriented along the Cu-O (Ni-O) bond direction and the
antiferromagnetic wavevector is at $Q_{AF} = (\pi,\pi)$.
In each plot, we integrate over an energy window of $\pm 0.2 J_a S$. %YAO S added
For vertical stripes, we show each plot for a twinned pattern of stripes,
adding the intensity from domains rotated $90^o$ with respect to each other.  

We first consider vertical stripes.  
Figs.~\ref{fig.vs4_0.4} and~\ref{fig.vs4_1.0} show results for vertical
site-centered stripes at coupling ratio $J_b/J_a = 0.4$ and $J_b/J_a=1.0$, respectively.
Figs.~\ref{fig.vb4_0.4} and~\ref{fig.vb4_1.5} show results for vertical
bond-centered stripes at coupling ratio $|J_b/J_a| = 0.4$ and  
$|J_b/J_a|=1.5$, %DY absolute value
 respectively.
 
The main incommensurate (IC) peaks indicating stripe order have high intensity at low 
energy in Figs.~\ref{fig.vs4_0.4}-\ref{fig.vb4_1.5}.   
However, the low energy satellite peaks are so weak as to be virtually unresolvable
in the site-centered case (Figs.~\ref{fig.vs4_0.4} and~\ref{fig.vs4_1.0}),
while they are visible but still weak in the bond-centered case (Figs.~\ref{fig.vb4_0.4} and~\ref{fig.vb4_1.5}).
At $\omega=0$, the ratio between the main incommensurate peaks and the satellite peaks is $34$ for VS4, and it is $5.9$ for VB4.  This is the maximum intensity ratio, for, {\em e.g.}, a square wave pattern
that follows Figs.~\ref{fig:allstripes} (a) and (b), in which every occupied site has the same spin moment.
With such a dramatic suppression of the satellite peaks for VS4,
and given the fact that the spin wave cone emanating from the satellite peaks in the site-centered case rapidly diminishes in intensity with increasing energy,\cite{spinprb1}  it is not surprising that the 
satellite cones are too faint to be seen in the lowest energies of Figs.~\ref{fig.vs4_0.4} and~\ref{fig.vs4_1.0}.
Rather than evidence against stripes, the absence of observed satellite peaks
may simply be due to insufficient experimental resolution, especially
in the case of site-centered stripes. 
Any envelope softer than a 
square wave pattern will  diminish the satellite peaks even further.\cite{wochner98a}

At slightly higher energy (the middle left panel in Figs.~\ref{fig.vs4_0.4}-\ref{fig.vb4_1.5}),
notice that the intensity on the spin wave cones can be different on the 
``inner branch," the side closest to ($\pi,\pi$), and the ``outer branch,"
the side farthest from ($\pi,\pi$).  In fact, in Fig.~\ref{fig.vb4_0.4}  (VB4 at $J_b/J_a=-0.4$),
the intensity is strongest on the inner branch.  On the other hand, at higher 
coupling ratios  (Figs.~\ref{fig.vs4_1.0} and~\ref{fig.vb4_1.5}),
the strongest intensity is on the {\em outer} branch.
The intensity ratio between inner and outer branches is a
function of the coupling ratio $|J_b/J_a|$, as shown in Fig.~\ref{fig:intensityratio}.
To obtain the figure, we have taken a cut through $k$-space 
perpendicular to the stripe direction, along $(k_x,\pi)$,
and restricted the plot to a low energy, $E=0.4 J_a S$~. %YAO  J_a S added
We divide the peak intensity at the inner branch of the spin wave cone by
the peak intensity of the outer branch of the spin wave cone to derive the
inner to outer branch intensity ratio in Fig.~\ref{fig:intensityratio}.
Based on a fit of the results in Fig.~\ref{fig:intensityratio},
we find the following functional form for the intensity ratio of inner to outer branches
for vertical stripes of spacing $d=4$:  
\begin{equation}
{S(\mathbf{k}_{\rm in},E_o) \over S(\mathbf{k}_{\rm out},E_o)} = a + {b \over {c + |J_b / J_a|^{\mu}}}~,
\label{eqn:intensityratio}
\end{equation}
where $\mathbf{k}_{\rm in}$ denotes the wavevector of the inner branch
of the spin wave cone on the $(\pi,\pi)$ side, and $\mathbf{k}_{\rm out}$ denotes the wavevector
of the outer branch away from $(\pi,\pi)$ at a particular energy $E_o$. 
For VS4  (site-centered stripes) at $E_o=0.4J_a S$ we find that $a=0.85$, $b=0.076$, $c=0$, and $\mu = 0.89$.  
For VB4 (bond-centered stripes) at $E_o=0.4J_a S$ we find that $a=0.84$, $b=0.18$, 
$c=-0.033$, and $\mu = 0.84$.
For both site- and bond-centered stripes, for small enough $|J_b/J_a|$,  the intensity
ratio is so dramatic that without sufficiently high resolution, the entire spin wave
cone {\em will not be resolvable}, and instead only ``legs of scattering" will
be visible.  Several neutron scattering experiments on the cuprates report this
kind of behavior,\cite{tranquada04a,hayden04,mookchg} and our calculations here indicate that the behavior is simply due to weak effective spin coupling across the charge stripes.  

For a range of coupling ratios, the acoustic band has a saddlepoint at $(\pi,\pi)$ which
has many similarities to the ``resonance peak" observed in the cuprates. 
These saddlepoints can be seen in the bottom right panel of Figs.~\ref{fig.vs4_0.4}-\ref{fig.vb4_1.5}.
Here, the intensity is gathered into one main peak at the antiferromagnetic wavevector $Q_{AF}$,
where the maximum intensity is higher than that at nearby energies.  
The saddlepoint structure gives rise to an hourglass shape emanating
from the resonance peak for twinned stripes in energy {\em vs} wavevector 
plots.\cite{uhrigsaddlepoint,mookchg,ybco6.7,ybco,normanhourglass} 
The resonance peak associated with a saddlepoint is more pronounced at low coupling ratio.  
For stronger coupling ratio, the weight in the spin wave cones is shifted to the outer branch, and
the extra intensity due to the saddlepoint at ($\pi,\pi$) is reduced.

%EC added this paragraph
For the case of vertical, site-centered stripes of spacing $p=4$ (VS4),  for $J_b=0.05J_a$, the resonance energy is $E_{\rm res} = 0.63J_a S$, and for $J_b = 0.2 J_a$, the saddlepoint is at $E_{\rm res} = 1.2 J_a S$.  
Assuming the value of $J_a$ ($\approx 140 {\rm meV}$\cite{coldea01})
is relatively unchanged upon doping, this gives a range of $E_{\rm res} = 44-84{\rm meV}$.
This encompasses the range $E_{\rm res} = 50-60{\rm meV}$ given in Ref.~\cite{tranquada04a} 
for the resonance peak in LBCO.  
For the case of vertical, bond-centered stripes of the same spacing (VB4),  with $J_b = 0.05J_a$, the saddlepoint is at $E_{\rm res} = 0.6J_a S$, and for $J_b = 0.2 J_a$, $E_{\rm res} = 1.1J_a S$.  This corresponds to  a range of $E_{\rm res} = 42-77{\rm meV}$, again encompassing the experimental range $E_{\rm res} = 50-60{\rm meV}$ in Ref.~\cite{tranquada04a}.   This would indicate that to match the resonance energy in LBCO requires that $0.05J_a \lesssim J_b \lesssim 0.2J_a$.\cite{spinprl}

Although doping is not explicitly in our model (rather we treat the stripe spacing in a phenomenological manner),  low energy neutron scattering indicates that the stripes move closer together as doping is 
%EC added reference
increased.\cite{yamada,mook2}
The energy of the saddlepoint resonance increases monotonically as the stripes are moved closer together in our model, holding $J_a$ and $J_b$ fixed.  
If $J_a$ and $J_b$ are relatively independent of doping (and we believe this is physically reasonable), then we would expect doping to increase the energy scale of, e.g., the resonance, as is observed in underdoped cuprates.  However, predicting how $J_a$ and $J_b$ depend on doping is beyond the scope of our present model.   
Another important piece of physics in the cuprates is that the intensity of the``resonance peak" increases
below $T_c$.  However, since we do not explicitly include superconductivity,
our model does not address the observed increase 
in intensity of the resonance peak as superconductivity onsets.

It is also possible to tune the coupling ratio $|J_b/J_a|$ so that the spin wave bands
cross instead of exhibiting a saddlepoint.  
This happens at $J_b/J_a = 1$ for site-centered stripes, and at $|J_b/J_a| = 0.56$
for bond-centered stripes.   It is unlikely that such a crossing would be observed
experimentally,\cite{balatsky} because of the fine-tuning it would require. 
The high energy structure of the spin wave response
therefore depends on whether the coupling ratio is above or below these 
special points.  For small coupling ratios, the generic behavior is that there is a high energy square-shaped continuum above the saddlepoint, with vertices rotated $45^o$ from the low energy incommensurate peaks,\cite{spinprl} 
reminiscent of the universal high energy behavior recently reported in
LBCO\cite{tranquada04a} and YBCO.\cite{hayden04}
For very weak coupling, the highly anisotropic spin wave structure
yields flat square edges in the high energy structure.  As the coupling ratio
$|J_b/J_a|$ is
increased, the square edges distort.  For higher coupling ratios, at and beyond the 
``crossing point", the high energy cross sections display a variety of patterns, including circular\cite{keimer04}
and square-shaped continua, as well as more complicated patterns.  
For large coupling ratios,
the high energy square-shaped continuum has vertices that are along the same
direction as the low energy peaks, as in Fig.~\ref{fig.vb4_1.5}.

%%%%%%%%%%%%%%%%%%%%%%%%%%%%%%%%%%
\section{Results: Spectra of Diagonal Stripes \label{results:diag}}
%%%%%%%%%%%%%%%%%%%%%%%%%%%%%%%%%%

%%%%%%%%%%%%%%% FIGURES:  Diagonal Cuts %%%%%%%%%%%%%%%%%%%
\begin{figure}
\resizebox*{1\columnwidth}{!}{\includegraphics{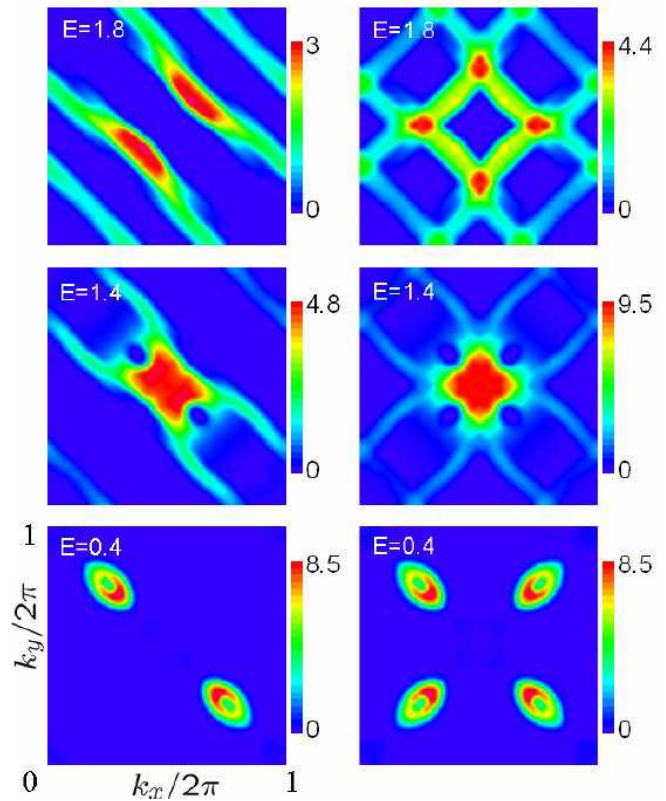}}
\caption{(Color online) DB2 at $J_b=-0.1J_a$: Constant energy cuts with windows of $0.2Ja$ for diagonal,
  bond-centered stripes of spacing $d=2$ with $J_b=-0.1J_a$. The left column is untwinned
  and the right one is twinned.   The energy $E$ is in units of $J_a S$.
  \label{cut.db2_0.1}}
\end{figure}
%%%%%%%%%%%%%%%%%%%%%%%%%%%%%%%%%%
%\begin{figure}
%\resizebox*{1\columnwidth}{!}{\includegraphics{DB2_2.eps}}
%\caption{Constant energy cuts with windows of $0.2Ja$ for diagonal,
%  bond-centered stripe DB2 with $J_b=-2.0 J_a$.   The energy $E$ is in units of $J_a S$.
%  The energies  for the figure are E=0.4, 1, 1.4, 1.5 , 1.8 and 2 arranged clockwise from lower left corner. 
%  \label{cut.db2_2.0}}
%\end{figure}
%%%%%%%%%%%%%%%%%%%%%%%%%%%%%%%%%
\begin{figure}
\resizebox*{0.95\columnwidth}{!}{\includegraphics{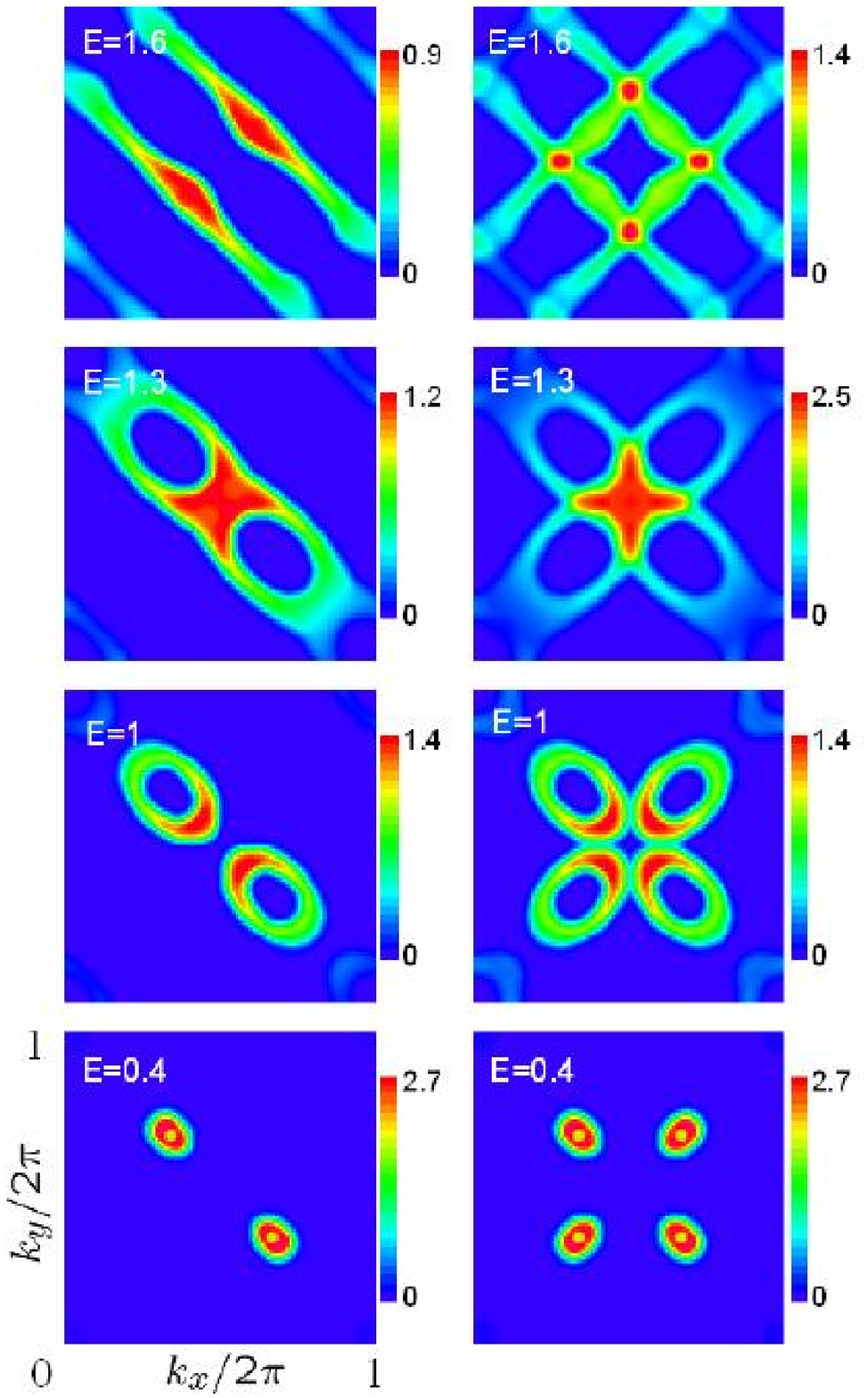}}
\caption{(Color online) DS3 at $J_b=0.1 J_a$: %DY  It's + sign 
 Constant energy cuts with windows of $0.2Ja$ for diagonal,
site-centered stripes of spacing $d=3$ at $J_b=0.1 J_a$.   The left column is untwinned
  and the right one is twinned.  The energy $E$ is in units of $J_a S$.
  \label{cut.ds3_0.1}}
\end{figure}
%%%%%%%%%%%%%%%%%%%%%%%%%%%%%%%%%%
\begin{figure}
\resizebox*{0.95\columnwidth}{!}{\includegraphics{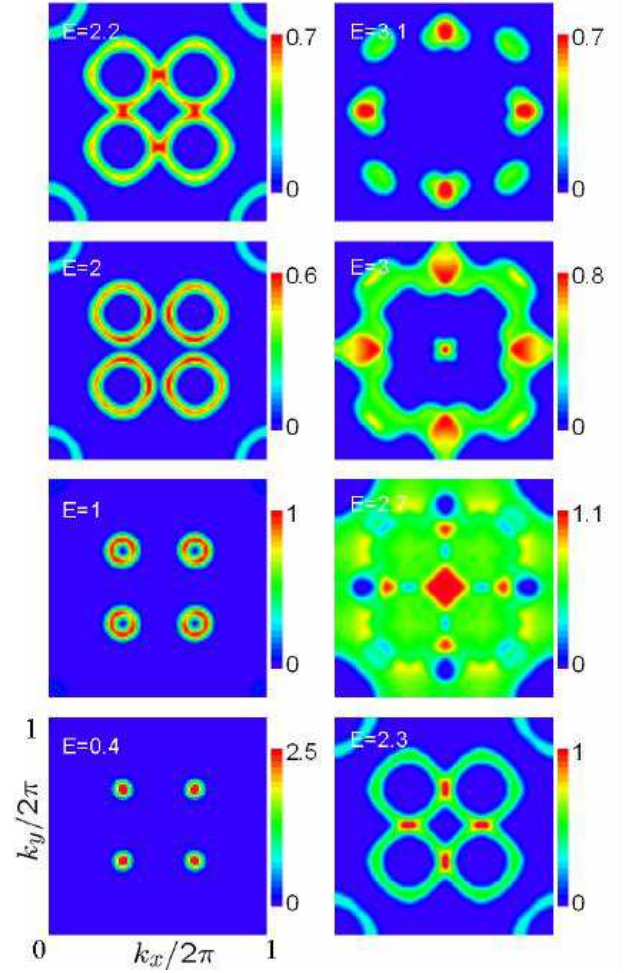}}
\caption{(Color online) DS3 at $J_b=0.5 J_a$: %DY It's + sign
 Constant energy cuts with windows of $0.2Ja$ for diagonal,
site-centered stripes of spacing $d=3$ at  $J_b=0.5 J_a$.   
Each panel is twinned.  
The energy $E$ is in units of $J_a S$.
%$The energies are E=0.4, 1, 2, 2.22.7 and 3.1 from bottom to top.$ %DY: not necessary 
  \label{cut.ds3_0.5}}
\end{figure}
%%%%%%%%%%%%%% FIGURE: Intensity Ratio %%%%%%%%%%%%%%%%%%%%
\begin{figure}[Htb]
{\centering
  \subfigure[DB2 Intensity Ratio]{\resizebox*{0.8\columnwidth}{!}{\includegraphics{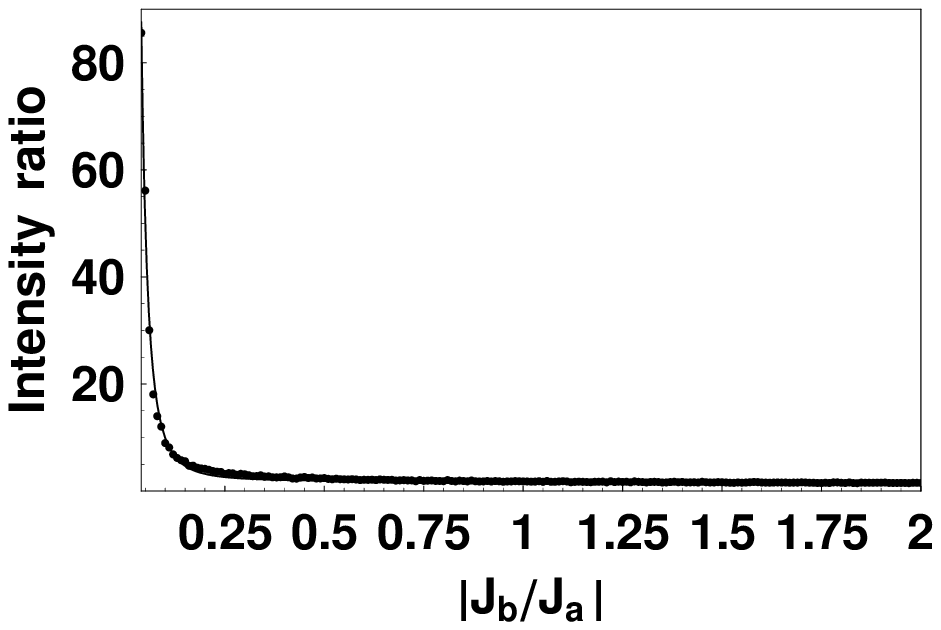}}}
  \subfigure[DS3 Intensity Ratio]{\resizebox*{0.8\columnwidth}{!}{\includegraphics{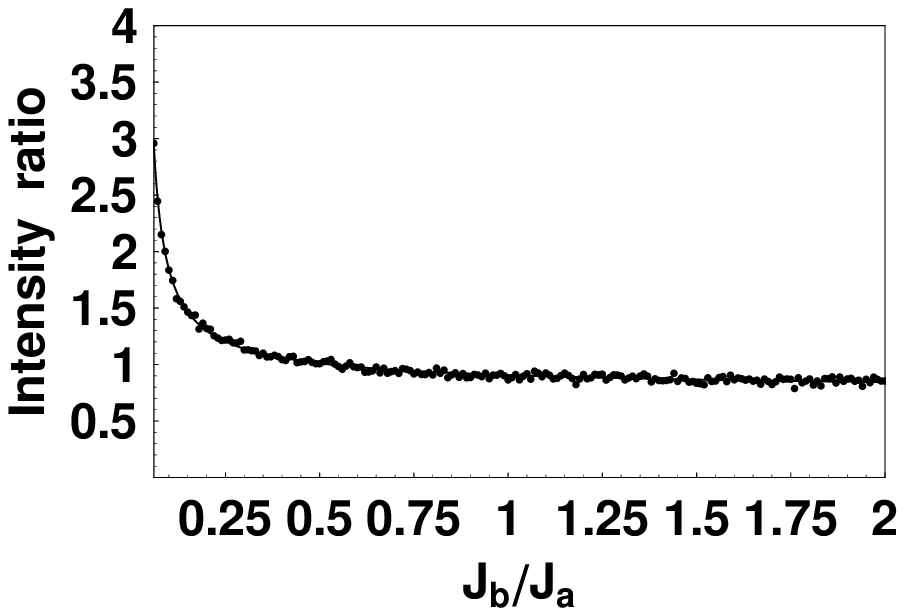}}} \par}
\caption{Intensity ratio between the ``inner" and ``outer" branches of the spin wave
cones as a function of coupling ratio $|J_b/J_a|$ for diagonal stripes. 
%DY Absolute value 
%The ratio is calculated at constant low energy $E=0.4J_aS$. %DY
 The 
solid points are our numerical calculations of the intensity ratio.  
The solid line is a fit to the results, as explained in the text.  
(a) DB2:  Intensity ratio for diagonal bond-centered stripes of spacing $d=2$
at $E=0.4J_a S$. 
(b) DS3:  Intensity ratio for diagonal site-centered stripes of spacing $d=3$
at $E=J_a S$.    
}
\label{fig:diag.intensityratio}
\end{figure}
%%%%%%%%%%%%%%%%%%%%%%%%%%%%%%%%%%%%%%%%

We now consider diagonal stripes.  Diagonal stripes have been observed  
in the nickelates,\cite{tranquadaLSNO,boothroyd} 
and at low doping in the cuprates, {\em e.g.}, for $x<0.05$ in 
La$_{2-x}$Sr$_x$CuO$_4$ 
(LSCO).\cite{wakimoto,lsco1d}
When the crystals are detwinned, the diagonal stripe incommensurate 
peaks are also untwinned in the cuprates.\cite{wakimoto,lsco1d}  To make a
comparison with the experiments on diagonal stripes, we report  results for untwinned stripes as well as twinned stripes.  

%On the other hand, the vertical stripes which are observed at higher doping show
%a twinned structure.  In lanthanum compounds, the twinning of vertical stripes
%is due to the tendency of stripes to run in orthogonal directions in 
%neighboring copper-oxygen planes.  In YBCO, where the Cu-O chains present an orienting
%potential, detwinning the crystal detwins the vertical stripes\cite{mook00,keimer04}.
%However, most data on YBCO is on twinned crystals.  For these reasons, we 
%report intensity plots of twinned vertical stripes.  

The expected neutron scattering intensities for diagonal stripes depend
starkly on whether the stripes have an even or odd spacing.  
Diagonal bond-centered stripes of odd spacing generally display net ferromagnetism.
This is because diagonal bond-centered domain walls have a net magnetic moment, 
and at odd domain wall spacing, each domain wall will have the same moment.
This dramatically changes the nature of the Goldstone modes from linear to quadratic in $|k-k_{o}|$.
For site- or bond-centered diagonal stripes of even spacing, the number of magnetic reciprocal lattice vectors 
doubles, and then even $Q_{AF}=(\pi,\pi)$ is a reciprocal lattice vector.
Since stripes are arrays of {\em antiphase} domain walls in the
antiferromagnetic texture, there can be no net antiferromagnetism,
and hence no zero frequency weight  at $Q_{AF}$.  Nevertheless,
a spin wave cone emanates from the $(\pi,\pi)$ point when the diagonal
spacing is even.  The cone has no weight at zero frequency, and gains a faint appearance
as energy is increased.\cite{spinprb1}  Because of this unique band structure, even-spaced diagonal
stripes cannot display a saddlepoint in the acoustic band at $(\pi,\pi)$.
There is, however,  a saddlepoint in the next optical band.

We show results for diagonal stripes in Figs.~\ref{cut.db2_0.1}-\ref{fig:diag.intensityratio}.
As with vertical stripes, the intensity profile of the spin wave cones at low energy is a 
function of $|J_b/J_a|$.
 For weak coupling $J_b$, the weight is strongly 
gathered near the $(\pi,\pi)$ point.
In Fig.~\ref{fig:diag.intensityratio}, we plot the intensity ratio between the 
inner and outer branches of the spin wave cones at small energy,
for both DB2 (diagonal bond-centered stripes of spacing $d=2$, Fig.~\ref{fig:diag.intensityratio}a) and
DS3 (diagonal site-centered stripes of spacing $d=3$, Fig.~\ref{fig:diag.intensityratio}b).  
To obtain the figures, we have taken a cut through $k$-space 
perpendicular to the stripe direction, along $(k_x,-k_x)$,
and we restrict the plots to a low energy, $E=0.4J_aS$.
We plot the peak intensity at the inner branch of the spin wave cone (toward $(\pi,\pi)$) divided by
the peak intensity of the outer branch of the spin wave cone (away from $(\pi,\pi)$).
By fitting  the results in Fig.~\ref{fig:diag.intensityratio} for diagonal stripes
we find that it is well described by the same functional
form that we used for vertical stripes, although the fitted constants are different.
Based on a fit to Eqn.~\ref{eqn:intensityratio},
we find that for DB2 at $E=0.4J_a S$ (Fig.~\ref{fig:diag.intensityratio}a),  $a = 1.8$, $b=0.022$, $c=0$, and $\mu = 2.6$. %DY
There is a slight deviation from the fit near $J_b = -0.25 J_a$. %YAO "-'' added
For DS3 at $E=J_a S$ (Fig.~\ref{fig:diag.intensityratio}b),  we find that $a = 0.74$, $b=0.15$, $c=-0.11$, and $\mu = 0.61$. 
As with vertical stripes, for diagonal stripes of both the site- and bond-centered type, 
at  small enough coupling ratio $|J_b/J_a|$, the intensity on the outer branch is so 
weak that only part of the spin wave cone will be visible without sufficient
experimental resolution, indicating that diagonal stripes can also display
``legs of scattering" when the spin coupling across the charge stripe is weak.  
(Similarly, large coupling across the charge stripe can display outwardly dispersing
``legs of scattering".)

In Fig.~\ref{cut.db2_0.1} we show constant energy cuts 
for DB2, a bond-centered diagonal stripe configuration with 
spacing $d=2$ between domain walls, at weak coupling across
the charge stripes, $J_b = -0.1 J_a$.
We report untwinned intensity plots in the left column, 
and twinned intensity plots in the right column.  
At low energies, the twinned intensity 
plots show four spin wave cones dispersing from the
incommensurate peaks, with weight concentrated near the $(\pi,\pi)$ point.

At higher energies in Fig.~\ref{cut.db2_0.1},
because the stripe spacing $d$ is even, 
the acoustic band cannot support a saddlepoint at $(\pi,\pi)$,
as discussed above.  
Rather, it is the optical band that has a saddlepoint, appearing 
at $E=1.4 J_a S$.  Like the resonance peaks in vertical stripes, the
saddlepoint\cite{gruninger} has higher intensity, with 
low energy and high energy branches emanating from it.
Due to the weak coupling across charge stripes
$|J_b| \ll J_a$,
%DY absolute value
the high energy branches emanating from the saddlepoint
are rather flat, giving rise to a high energy square shaped continuum in the twinned plots, rotated
$45^o$ from the low energy peaks.  

In Fig.~\ref{cut.ds3_0.1},
we show results for DS3, diagonal site-centered stripes of spacing
$d=3$, with $J_b = 0.1 J_a$.  Because the coupling $J_b$ is weak,
the spin wave cones have weight gathered near $(\pi,\pi)$. 
The low energy dispersion for DS3 at this coupling is much steeper than that for
DB2 at the same coupling strength, 
because unlike DB2, DS3 has no magnetic reciprocal lattice vectors at $(\pi,\pi)$,
and so the spin wave cones of DS3
continue up in energy to a saddlepoint at $(\pi,\pi)$.  
The saddlepoint occurs at $E=1.3 J_a S$, and it has extra intensity
when twinned.  As with other saddlepoints at low $J_b$, 
there are high energy and low energy branches emanating from it,
resulting in an hourglass shape in $E$ {\em vs.} $k$ for twinned stripes. 
The weak coupling $J_b \ll J_a$ gives rise to a square shaped continuum above the saddlepoint, with vertices rotated $45^o$ from the low energy peaks, a familiar pattern at high energy.

In Fig.~\ref{cut.ds3_0.5}, we show DS3 at higher coupling ratio, $J_b = 0.5J_a$.
This pattern and coupling has been shown to capture many essential features\footnote{There is
an as-yet unexplained extra intensity which appears above $20$meV, which
was modeled phenomenologically with the addition of a gapped spectrum in Ref.~\cite{tranquadaLSNO}.}
of the data for the nickelate compound La$_{2-x}$Sr$_x$NiO$_4$ at $x=1/3$, and to some extent $x=0.275$ as well.\cite{tranquadaLSNO}
To more fully compare with this experiment, we show results only for twinned stripes.
Our results may be compared with Figs.~$2,~5,$ and $10$ of Ref.~\cite{tranquadaLSNO}.
(However note that the axes of the plots in Ref.~\cite{tranquadaLSNO}
are rotated $45^o$ from our plots.)
At low energy (up to $E=J_a S$), the spin wave cones have intensity
peaked on the outer branch.  However, at higher energy,
$E=2J_a S$, the intensity has shifted, and is now peaked on the {\em inner} branch.
The spin wave cones remain remarkably circular through this energy range, despite the 
anisotropic coupling ratio. 
As the spin wave cones touch at $E=2.3 J_a S$, there are four regions
of high scattering radiating out from $(\pi,\pi)$.  
As the cones merge, they form a central peak at $(\pi,\pi)$,
surrounded by four smaller regions of high scattering, rotated $45^o$ from
the low energy peaks.  These four smaller peaks are mainly due to the addition
of two twinned stripe patterns.  At higher energy, these four peaks move outward
(occupying the corners of the graph in Fig.~$5$(f) of Ref.~\cite{tranquadaLSNO}),
while the peak at $(\pi,\pi)$ diminishes.  Finally at higher energy $E=3.1
J_a S$, 
the peak at $(\pi,\pi)$ is no longer visible.  This indicates that rather than being a 
saddlepoint, as happened for $J_b \ll J_a$, by the time the coupling ratio has
reached $J_b = 0.5J_a$, there is no longer a saddlepoint at $(\pi,\pi)$.  Rather,
there is simply a band edge at $E=3J_a S$.  This is also consistent with the fact that
the $(\pi,\pi)$ peak in this case does not display significantly higher intensity compared to
that of nearby energy cuts.

%%%%%%%%%%%%%%%%%%%%%%%%%%%%%%%%%%
\section{RESULTS: Spectra of Checkerboards \label{sec:check}}

%%%%%%%%%%%%%%%%%%%%%%%%%%%%%%%%%%
\begin{figure}
{\centering 
%\resizebox*{0.95\columnwidth}{!}{\includegraphics{cs3_0.1.eps}} \par}
\resizebox*{0.99\columnwidth}{!}{\includegraphics{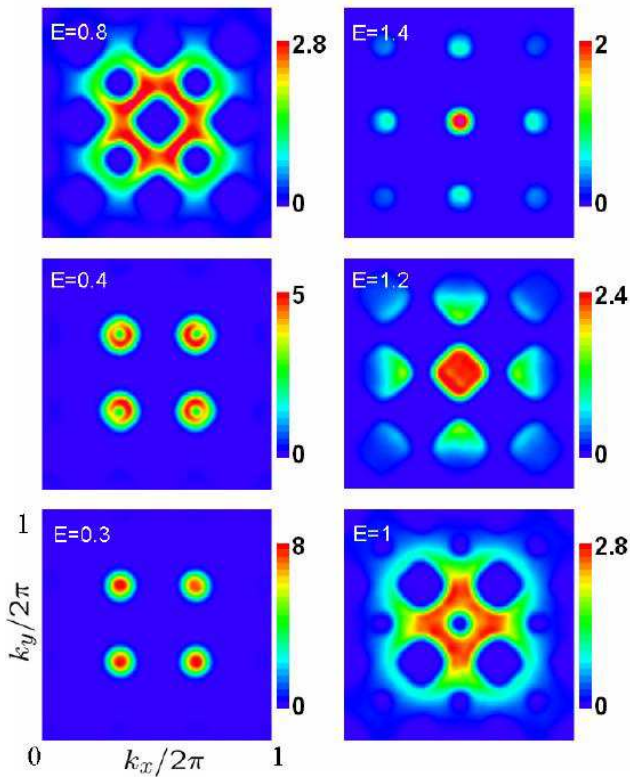}} \par}
\caption{(Color online) CS3 at $J_b=0.1J_a$: Constant energy cuts for site-centered checkerboard pattern with stripe spacing
  $d=3$ at $J_b=0.1J_a$.  The energy $E$ is in units of $J_a S$. 
\label{fig:cs3}}
\end{figure}
%%%%%%%%%%%%%%%%%%%%%%%%%%%%%%%%%%

We show in Fig.~\ref{fig:cs3} the expected scattering response from a typical 
simple checkerboard pattern.  We show results for a site-centered checkerboard pattern
of spacing $d=3$,  the real space pattern of which is shown in 
Fig.~\ref{fig:checkerbond}(a).
At low energies, simple checkerboards
which use the charged lines as antiphase domain walls in the spin texture
yield zero frequency peaks (and therefore low energy spin wave cones) which
are {\em in the  wrong direction}.  That is to say, for vertically placed domain
walls (as required by both the charge incommensurate peaks in neutron scattering
on LNSCO at dopings $x>0.05$\cite{tranqnature,ichikawa} and the 
STM Fourier transform peaks in BSCCO\cite{checknccoc,aharon,hoffman}) the spin incommensurate peaks
are rotated $45^o$ from those observed experimentally.\cite{vojta-sachdev}  
Simple checkerboard 
pattens like these always give IC spin peaks rotated $45^o$ from the
%EC added reference 
direction of the charge IC peaks,\cite{vojta-sachdev,andersen03}
contrary to what has been 
seen experimentally. 
However, more complicated checkerboard patterns\cite{abbamonte}
may be capable of fitting the low energy data, 
and we plan to explore the finite frequency response of these in a future publication.\cite{future}

Simple bond-centered checkerboards of odd spacing ($d \in odd$)
suffer yet another ill:  they support a net magnetization as shown in Fig.~\ref{fig:checkerbond}(b).
As a result, they display a  ferromagnetic spin peak at $Q_{F} = (0,0)$,
and spin waves which are quadratic rather than linear in $|k-k_o|$,
where $k_o$ is a magnetic reciprocal lattice vector.
Net ferromagnetism is not attainable for simple vertical stripes, 
but it is possible with bond-centered diagonal stripes of odd spacing, 
as discussed in Sec.~\ref{results:diag}.

The spin wave cones at low energy in Fig.~\ref{fig:cs3}
have weight gathered on the inner branches
on the side nearest $(\pi,\pi)$.  The intensity in these cones merges into a square-like
pattern as energy is increased, before the band ends with a peak at 
$E=1.4J_aS$ and $(\pi,\pi)$.
The high energy part of the acoustic band also has incommensurate peaks
which are in the {\em correct} direction for the low energy IC spin peaks,
but are overwhelmed in intensity by the central peak at $Q_{AF}$.  
The high energy peak at $Q_{AF}$ which marks the top of the acoustic band
is unlike the resonance peak observed in the experiment, since there is no scattering signal emanating from it at higher energy.\cite{vojta-sachdev}
Rather, above $E=1.4 J_a S$ in Fig.~\ref{fig:cs3}, there
is a spin wave gap to a rather flat optical band.

The form of the energy in this checkerboard configuration (CS3) may
be calculated analytically.  
There are four spins in the unit cell, as shown in Fig.~\ref{fig:checkerbond}(a),
and  only two bands.  
Along the  direction diagonal to the domain walls $(k_x,k_x)$,
the dispersions for the accoustic and
optical bands are   

\begin{eqnarray}
\frac{\omega_{acc}}{J_a S}= 4\sqrt{\lambda}|\sin{\frac{3k_x}{2}}|,\\
\frac{\omega_{op}}{J_a S}=2(1+\lambda)~,
\end{eqnarray} 
with $\lambda=J_b/J_a$. Notice that the optical band is flat in this direction.
Along the direction parallel to the domain walls $(k_x,0)$,
the dispersions are 
\begin{eqnarray}
\big{(}\frac{\omega}{J_a S}\big{)}^2/2&=&(1+3\lambda+\lambda^2-\lambda \cos{3k_x}) \\ \nonumber
&\pm& (1+\lambda)\sqrt{1+\lambda^2+2\lambda \cos{3k_x}}~,
\end{eqnarray}
where  the $-$ sign refers to the
acoustic band, and the $+$ sign refers to the optical band.  
The spin wave velocities in the diagonal and parallel directions are
\begin{eqnarray}
v_{diag}=\frac{3\sqrt{\lambda}}{2} v_{AF},\\
v_{||}=\frac{3\sqrt{\lambda}}{2\sqrt{2}}v_{AF}~.
\end{eqnarray}

%%%%%%%%%% CHECKERBOARD FIGURES %%%%%%%%%
\begin{figure}
{\centering \resizebox*{0.95\columnwidth}{!}{\includegraphics{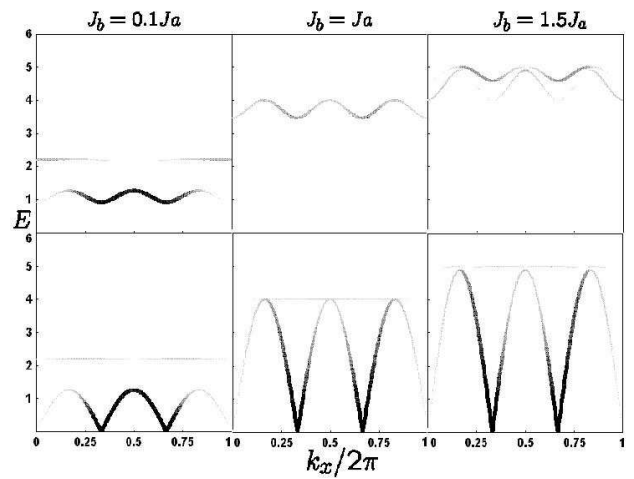}} \par}
\caption{CS3 at $J_b=0.1J_a$:  Dispersion and intensities for a site-centered checkerboard pattern with stripe spacing $d=3$ at $J_b=0.1J_a$. 
The upper pannel is along ($k_x,\pi$) direction ; the lower pannel
corresponds to the diagonal direction ($k_x,k_x$).  The energy $E$ is in units of $J_a S$. \label{spec_cs3}}\end{figure}
%%%%%%%%%%%%%%%%%%%%%%%%%%%%%%%%%%

These band structures (with numerically calculated intensities)
are shown in Fig.~\ref{spec_cs3} in the $(k_x, \pi)$ and $(k_x,k_x)$ directions.
There are always two bands present, although one is often quite weak
compared to the other.  
The coupling at which the two bands touch each other is 
$J_b = J_a$.
Fig.~\ref{spec_cs3}(a) shows the gap that is
present between the acoustic and optical bands, and that
there is no high energy structure emanating from the 
peak at $(\pi,\pi)$ which terminates the acoustic band.  

%%%%%%%%%%%%%%%%%%%%%%%%%%%%%%%%%%
\section{Conclusions}
%%%%%%%%%%%%%%%%%%%%%%%%%%%%%%%%%%
In conclusion, we have studied the expected inelastic neutron scattering intensities 
from spin waves
for a variety of spin-ordered phases, including vertical and diagonal stripes
and a simple checkerboard pattern.  We find that the inelastic response
is very sensitive to the coupling across domain walls
throughout the energy ranges studied.    In addition, we wish to emphasize
that the {\em elastic} response can also hold surprises.
For example, we find for vertical stripes of spacing $d=4$
that when the domain walls are site-centered (VS4), the ratio between
the main IC peaks and the next satellite peak is at least $34$.
This number is independent of coupling strength, and it holds for the most sharply spin-ordered case
of a square-wave profile to the antiphase domain walls.
Stripes in a real material are expected to have a softer profile due to
quantum effects neglected here, and this will diminish the satellite peaks even more.
For the bond-centered $d=4$ case (VB4), the satellite peaks are easier to detect,
with a ratio of at least $5.9$ between the main IC peaks and the
next satellite peaks, even for the most extreme case of a square-wave profile.  
However, this is still below the noise for current experimental resolution.

For vertical stripes at finite but low energies, the intensity around the spin wave cones is rarely uniform.
Rather, it is shifted 
either toward or away from the $(\pi,\pi)$ point by an amount which depends 
on the  coupling ratio
$|J_b/J_a|$.  For weak coupling across the charge stripes, {\em e.g.} $|J_b| < J_a$, 
the weight is gathered on the inner branch toward the $(\pi,\pi)$ peak,
indicating that weak stripe coupling may be responsible for the observed
``legs" of scattering in stripe-ordered cuprates at low frequency. 
For stronger coupling across the charge stripes, {\em e.g.} $|J_b| > J_a$, the intensity
shifts to the outer branch.  In either case, without sufficient experimental resolution, only part of
the spin wave cones will be visible at low energy.  

At intermediate energies and for weak coupling, the acoustic band
displays a saddlepoint whose intensity profile mimics that of the 
``resonance peak" observed in cuprate superconductors.  For twinned
stripes, a saddlepoint displays the characteristic hourglass  shape in 
$E$ {\em vs.} $k$ plots seen in some experiments, with both low and high energy legs emanating from 
a resonance-like peak.\cite{mookchg,ybco6.7,ybco,normanhourglass,uhrigsaddlepoint}

At high energies and weak coupling $J_b$, vertical stripes show a square-shaped continuum 
which is rotated $45^o$ from the low energy IC peaks.  As we have shown
previously,\cite{spinprl} very weak coupling
captures the spin wave excitations of stripe-ordered LBCO
at all measured energies, and at higher energies it bears a striking resemblance 
to the high energy excitations of YBCO.  The gap to spin $S=1$ excitations in YBCO 
is likely due to the stripes being too weakly coupled, and therefore quantum disordered.
However, at high energy, the quantum critical excitations\cite{vojta-ulbricht,uhrig04a,tranquada04a,seibold04}
strongly resemble the spin waves 
studied here, which may be due to the proximity of a QCP with
small critical exponent $\eta = 0.037$.\cite{spinprl}
%EC added the next 2 sentences.
Near the QCP, one way to distinguish semiclassical spin waves from quantum critical excitations
is through the lineshapes.   Whereas semiclassical spin waves produce a Lorentzian lineshape, quantum criticality yields a power law cusp.  
For intermediate and larger couplings $J_b$, the high energy response can have  a variety of
shapes.  We have shown here that circular continua are possible, as well as 
square-shaped continua which either have the same orientation as the low energy peaks, 
or are rotated $45^o$ from that direction.

For diagonal stripes, as with vertical stripes,  the low energy spin wave cones have intensity
profiles which depend on the strength of the coupling between the stripes.
For weak coupling, the intensity is peaked on the inner branches, while for strong coupling,
it is peaked on the outer branches.  In either extreme case, the entire spin-wave
cone may not be visible in an experiment without sufficient resolution.  
At intermediate energy, weakly coupled diagonal stripes also have a 
saddlepoint in the acoustic band, with properties akin to the resonance peak.
At high energy, weakly coupled diagonal stripes display a high energy square-shaped
continuum, rotated $45^o$ from the low energy peaks.  
At higher coupling ratio $|J_b/J_a|$, the saddlepoint in the acoustic band
at $(\pi,\pi)$ is lost, and a variety of high energy scattering patterns are possible.  
Diagonal bond-centered stripes of odd spacing display net ferromagnetism,
changing the nature of Goldstone modes from linear to quadratic in $|k-k_o|$.
Both site- and bond-centered stripes of even spacing have magnetic reciprocal lattice
vectors which include $Q_{\rm AF} = (\pi,\pi)$.  Because of the antiphase nature
of the domain walls, zero frequency is forbidden at this peak, but the 
(very low intensity) spin wave cone emanating from $(\pi,\pi)$ precludes
a saddlepoint at $Q_{\rm AF}$ in the acoustic band.  

For intermediate coupling $J_b = 0.5 J_a$, 
 the expected scattering intensity
from diagonal, site-centered stripes of spacing $d=3$ strongly resembles that in
La$_{2-x}$Sr$_x$NiO$_4$ at $x=1/3$ and $x=0.275$, as pointed out in Ref.~\cite{tranquadaLSNO}.  
We find for this configuration
that while the intensity is peaked on the outer branch of the spin wave cone
at low energy, it moves toward the inner branch as energy is increased.  
Furthermore, the spin wave cones are remarkably circular, despite the
anisotropic coupling ratio.  With increasing energy, the spin wave
cones merge, eventually gathering weight at the central peak $(\pi,\pi)$.
However, rather than a saddlepoint, this coupling  has a band edge at $(\pi,\pi)$.  

Simple checkerboards, on the other hand, have low energy spin IC peaks
rotated $45^o$ from the observed spin peaks in neutron scattering.\cite{vojta-sachdev}
In addition, for weak coupling
across the domain walls $|J_b| \ll J_a$, 
rather than showing a resonance-like saddlepoint,
the acoustic band has a band edge at $(\pi,\pi)$,\cite{vojta-sachdev} 
and therefore no branches emanating from it at high energy.  Although the simple checkerboard
studied here is incompatible with the experimental data, 
this does not rule out the possibility of 
more complicated checkerboard patterns\cite{abbamonte}
which may be able to capture the correct orientation of the low energy
spin and charge IC peaks.

%%%%%%%%%%%%%%%%%%%%%%%%%%%%%%%%%%
\section*{Acknowledgments}
%%%%%%%%%%%%%%%%%%%%%%%%%%%%%%%%%%
It is a  pleasure to thank Y.~S.~Lee, M.~Granath, A.~Sandvik, Y.~Siddis, P. Abbamonte, and J.~Tranquada for helpful discussions. 
This work was supported by Boston University (D.X.Y. and D.K.C.), and by the Purdue Research Foundation (E.W.C.).   

%\bibliography{bigbib}
\bibliographystyle{forprb}

\end{document}